\documentclass[11pt,english]{article}
\pdfoutput=1
\usepackage[T1]{fontenc}
\usepackage[latin9]{inputenc}
\usepackage{geometry}
\usepackage{babel}
\usepackage{array}
\usepackage{multirow}
\usepackage{amsmath}
\usepackage{amssymb}
\usepackage{graphicx}
\usepackage{esint}
\usepackage[unicode=true,pdfusetitle,
 bookmarks=true,bookmarksnumbered=false,bookmarksopen=false,
 breaklinks=false,pdfborder={0 0 1},backref=false,colorlinks=false]
 {hyperref}

\makeatletter

\providecommand{\tabularnewline}{\\}
\newcommand{\lyxdot}{.}

\numberwithin{equation}{section}
\numberwithin{figure}{section}
\numberwithin{table}{section}

\@ifundefined{showcaptionsetup}{}{%
 \PassOptionsToPackage{caption=false}{subfig}}
\usepackage{subfig}
\makeatother

\begin{document}

\title{Confinement in the $q$-state Potts model: an RG-TCSA study}

\author{M. Lencsés$^{1,2}$ and G. Takács$^{1,3}$\\
 ~\\
 $^{1}$MTA-BME \textquotedbl{}Momentum\textquotedbl{} Statistical
Field Theory Research Group\\
1111 Budapest, Budafoki út 8, Hungary\\
~\\
 $^{2}$Department of Theoretical Physics, Eötvös University\\
 1117 Budapest, Pázmány Péter sétány 1/A, Hungary\\
 ~\\
 $^{3}$Department of Theoretical Physics, \\
 Budapest University of Technology and Economics\\
1111 Budapest, Budafoki út 8, Hungary}

\date{23rd June 2015}
\maketitle
\begin{abstract}
In the ferromagnetic phase of the $q$-state Potts model, switching
on an external magnetic field induces confinement of the domain wall
excitations. For the Ising model ($q=2$) the spectrum consists of
kink-antikink states which are the analogues of mesonic states in
QCD, while for $q=3$, depending on the sign of the field, the spectrum
may also contain three-kink bound states which are the analogues of
the baryons. In recent years the resulting ``hadron'' spectrum was
described using several different approaches, such as quantum mechanics
in the confining linear potential, WKB methods and also the Bethe-Salpeter
equation. Here we compare the available predictions to numerical results
from renormalization group improved truncated conformal space approach
(RG-TCSA). While mesonic states in the Ising model have already been
considered in a different truncated Hamiltonian approach, this is
the first time that a precision numerical study is performed for the
$3$-state Potts model. We find that the semiclassical approach provides
a very accurate description for the mesonic spectrum in all the parameter
regime for weak magnetic field, while the low-energy expansion from
the Bethe-Salpeter equation is only valid for very weak fields
where it gives a slight improvement over the semiclassical results.
In addition, we confirm the validity of the recent predictions for
the baryon spectrum obtained from solving the quantum mechanical three-body
problem.
\end{abstract}

\section{Introduction\label{sec:Introduction}}

In this work we consider the scaling field theory corresponding to
the $q$-state Potts model in the vicinity of the critical point separating
the paramagnetic and ferromagnetic phases. In the absence of external
magnetic field, the field theory in the ferromagnetic phase has a
spectrum consisting of kinks (domain wall excitations) and is integrable
with a known factorized $S$ matrix \cite{ChimZam1992}. Switching
on a weak external magnetic field induces a linear potential between
the kinks, leading to their confinement. 

In the case of the Ising model ($q=2$) this scenario was proposed in \cite{McCoyWu1978},
and has been investigated using several methods since then. One approach
consists of a numerical Hamiltonian truncation method \cite{FonZam2003},
which allows the numerical determination of the resulting meson spectrum.
On the other hand, there are several approaches allowing predictions
for the spectrum, with the first prediction already made in \cite{McCoyWu1978},
which is in fact equivalent to finding the spectrum of quantum mechanical
bound states in a linear potential. Form factor perturbation theory 
which treats the strength of integrability breaking (such as introduced by 
the magnetic field)
as a small parameter can also be used to establish 
confinement \cite{Delfino:1996xp,Delfino:1997ya}.
The meson spectrum can also be computed using the WKB method \cite{FonZam2006},
and a low-energy expansion can be obtained from the Bethe-Salpeter equation \cite{FonZam2003,FonZam2006,Rutkevich2005,Rutkevich2009,Rutkevich2010}.
The validity of these approaches and their consistency has been established
not only on the qualitative, but also at a quantitative level to high
precision. 

In the $q=3$ case the spectrum is expected to be richer and contain
baryonic three-kink bound states besides the mesonic kink-antikink
ones \cite{Delfino2008265}. For the $3$-state Potts model, the Hamiltonian
truncation method used in \cite{FonZam2003,FonZam2006} is not available,
since it requires the knowledge of exact finite volume form factors
of the magnetization operator in the off-critical theory in zero magnetic
field. However, an alternative is provided by the truncated conformal
space approach (TCSA) introduced in \cite{YurZam1990}. Since construction
of low-energy spectrum in the TCSA does not depend on the assumption
of integrability, nor on the existence of small parameters, it can
be readily applied to non-integrable models,
as performed for the $q=2$ and $3$ cases in \cite{Pozsgay:2006wb} and \cite{LepTGZSDelf2009}, 
respectively;
for more recent applications to other non-integrable quantum field theories 
cf. \cite{Coser:2014lla,2015arXiv150503860K}.
In the case of the $3$-state Potts model, a previous study has already
confirmed qualitatively the expected mesonic and baryonic spectra
\cite{LepTGZSDelf2009}. 

More recently precise quantum mechanical and semiclassical predictions
for the mass spectrum have been obtained \cite{RutPottsMes2010,Rut2015},
which makes it possible to investigate the spectra in more detail.
For this purpose a more advanced implementation of TCSA is needed,
which is based on the renormalization
group improvements recently introduced and developed in \cite{FeveratiGrahamPearceTGZSWatts2008,konik2009,GiokasWatts2011,LencsesTakacs2014,Rychkov2015a,Rychkov2015b}.
In this work we discuss the application of this RG-TCSA method to the 
Ising and 3-state Potts models and
compare it to the theoretical predictions.

The outline of the present paper is as follows. In Section \ref{sec:Confinement-in--state}
we review the known results about confinement in the scaling $q$-state
Potts field theory. In Section \ref{sec:TCSA:-RG-and} we briefly
describe our TCSA implementation. Section \ref{sec:Results} contains
our results, starting with the Ising case as a testing ground
for our numerical procedures and then turning to the $3$-state Potts
model. In Section \ref{sec:Conclusion} we draw our conclusions.

\section{Confinement in $q$-state Potts field theory\label{sec:Confinement-in--state}}

\subsection{Brief overview\label{sub:Brief-overview}}

In this section we briefly overview the phenomenology of the $q$-state
model based on \cite{Delfino2008265}. The $q$-state Potts field
theory is the scaling field theory of the $q$-state Potts model which
is the generalization of the Ising model with $q$ different values
(colours) of the lattice variables \cite{Potts1952}. The lattice
Hamiltonian can be written as
\begin{equation}
\mathcal{H}=-\frac{1}{T}\sum_{\left\langle x,y\right\rangle }\delta_{s\left(x\right),s\left(y\right)}-H\sum_{x}\delta_{s\left(x\right),q}\label{eq:lattHam}
\end{equation}
The first term is the standard nearest-neighbour interaction and second
term is a generalized magnetic field in the ``direction'' of the
$q$-th colour. Without magnetic field ($H=0$) the theory has $S_{q}$
permutation symmetry and has a critical temperature $T=T_{c}$, below
which the system is in the ordered (ferromagnetic) phase while above
$T_{c}$ the system is in the disordered (paramagnetic) phase.

In this work we consider the ferromagnetic phase, in which there are
$q$ degenerate ground states (with all sites having the same colour),
and the elementary excitations are kinks corresponding to domain walls.
The action of the scaling field theory of the system can be written
as a perturbation of the conformal field theory (CFT) corresponding
to the critical point: 
\begin{equation}
S=S_{CFT}^{\left(q\right)}+\tau\int d^{2}x\varepsilon\left(x\right)+h\int d^{2}x\sigma\left(x\right)\label{eq:SFTaction}
\end{equation}
where the couplings $\tau$ and $h$ are related to the lattice couplings
\begin{eqnarray}
\tau & \sim & T-T_{c}\\
h & \sim & H
\end{eqnarray}
and $x=\left(x_{1},x_{2}\right)$ are Euclidean time and space coordinates.
The corresponding CFT \cite{Belavin:1984vu} is defined for $q\leq4$
and has the central charge \cite{Dotsenko1984312} 
\begin{equation}
c\left(q\right)=1-\frac{6}{t\left(t+1\right)}\label{eq:CentralCharge}
\end{equation}
where the parameter $t$ is related to $q$ via the relation
\begin{equation}
\sqrt{q}=2\ \sin\frac{\pi\left(t-1\right)}{2\left(t+1\right)}
\end{equation}
and the thermal and magnetic fields $\varepsilon$ and $\sigma$ are
identified with spinless relevant primaries $\Phi_{2,1}$ and $\Phi_{\left(t-1\right)/2,\left(t+1\right)/2}$
of the CFT and have scaling dimensions \cite{Dotsenko1984312,Nienhuis1984}
\begin{eqnarray}
2h_{\varepsilon}^{\left(q\right)}=X_{\varepsilon}^{\left(q\right)} & = & \frac{1}{2}\left(1+\frac{3}{t}\right)\\
2h_{\sigma}^{\left(q\right)}=X_{\sigma}^{\left(q\right)} & = & \frac{\left(t-1\right)\left(t+3\right)}{8t\left(t+1\right)}
\end{eqnarray}
The thermal operator preserves the $S_{q}$ symmetry while adding
the magnetic field breaks the permutation symmetry according to $S_{q}\rightarrow S_{q-1}$.

In absence of the magnetic field the theory is integrable and the
complete scattering theory is known \cite{Koberle1979209,Zamolodchikov1988,ChimZam1992}.
In the ferromagnetic phase there exist $q$ degenerate ground states
and the elementary excitations are kinks interpolating between them,
with their mass $m$ related to the coupling $\tau$ via the mass
gap relation \cite{Fateev199445} 
\[
\tau=\kappa^{(q)}m^{2-2h_{\varepsilon}^{(q)}}
\]
where 
\begin{eqnarray}
\kappa^{\left(2\right)} & = & \frac{1}{2\pi}\\
\kappa^{\left(3\right)} & = & \frac{\Gamma\left(\frac{3}{10}\right)\left[\Gamma\left(\frac{2}{3}\right)\Gamma\left(\frac{5}{6}\right)\right]^{6/5}}{4\times2^{1/5}\pi^{8/5}\Gamma\left(\frac{7}{10}\right)}\sqrt{\frac{\Gamma\left(-\frac{1}{5}\right)\Gamma\left(\frac{7}{5}\right)}{\Gamma\left(-\frac{2}{5}\right)\Gamma\left(\frac{6}{5}\right)}}=0.1643033\dots
\end{eqnarray}

In a magnetic field $h$ in the direction of the $q$-th colour the
degeneracy between the $q$ ground states is lifted. For $h<0$ there
is a single true vacuum and $q-1$ metastable ones, while for $h>0$
their roles are exchanged. The expectation values of the magnetic
field operator in the direction of the $\gamma$-th colour is in the
$\alpha$-th ground state of the zero field theory is written as
\begin{equation}
\left\langle \sigma_{\gamma}\right\rangle _{\alpha}\equiv\left\langle 0_{\alpha}\left|\sigma_{\gamma}\left(x\right)\right|0_{\alpha}\right\rangle =\frac{v^{(q)}}{q-1}\left(q\delta_{\gamma,\alpha}-1\right)
\end{equation}
where $v^{(q)}$ can be calculated using the formulas in \cite{Fateev1998652}:
\begin{eqnarray}
v^{\left(2\right)} & = & 1.3578383417\dots\times m^{1/8}\\
v^{\left(3\right)} & = & 1.9382577836\dots\times m^{2/15}
\end{eqnarray}
As a result, the difference of the energy density between the false
and the true vacua is given by \cite{Delfino2008265} 
\begin{equation}
\Delta\varepsilon=\delta\varepsilon_{\alpha}-\delta\varepsilon_{q}\simeq h\left(\left\langle \sigma_{q}\right\rangle _{\alpha}-\left\langle \sigma_{q}\right\rangle _{q}\right)=-\frac{v^{(q)}q}{q-1}h\qquad\alpha\neq q\label{eq:fv_endensity}
\end{equation}
The combination $v^{\left(q\right)}q/\left(q-1\right)$ is denoted
by $\beta^{\left(q\right)}$ and called the string tension; $\beta^{(q)}|h|$
gives the slope of the linear potential induced by the magnetic field.

\subsection{Mesonic and baryonic mass estimations\label{sub:Mesonic-and-baryonic}}

Here we present the various known estimations for the masses in the
confinement spectrum for the case of the Ising and the three state
Potts model.

\subsubsection{Meson masses in the Ising model\label{par:Meson-masses-in}}

The first mass estimation is the one obtained by McCoy and Wu \cite{McCoyWu1978}
\begin{equation}
m_{n}^{Ai}=m(2+\lambda^{2/3}z_{n})\label{eq:Airy_ising}
\end{equation}
where $-z_{n}$ is the $n$th zero of the Airy function and $\lambda$
is the dimensionless ratio
\begin{equation}
\lambda=\frac{\beta^{\left(2\right)}|h|}{m^{2}}
\end{equation}
This solution can be derived from the quantum mechanical system of
two kinks in a linear potential. The quantum mechanical system also
allows solutions corresponding to the zeros for the derivative of
the Airy function; however, the corresponding wave-functions are symmetric
and so forbidden due to the fermionic nature of the kinks. 

The WKB mass spectrum can be obtained by solving the quantization
condition 
\begin{eqnarray}
\frac{\sinh\left(2\vartheta_{n}\right)-2\vartheta_{n}}{\lambda} & = & 2\pi\left(n-1/4\right)\nonumber \\
m_{n}^{WKB} & = & 2m\cosh\left(\vartheta_{n}\right)\label{eq:wkb_ising}
\end{eqnarray}
It can be improved further by adding higher corrections in $\lambda$:
\begin{eqnarray}
\sinh\left(2\vartheta_{n}\right)-2\vartheta_{n} & = & 2\pi\left(n-1/4\right)\lambda+\sum_{k=1}^{\infty}\lambda^{k+1}S_{k}\left(\vartheta_{n}\right)\nonumber \\
m_{n}^{iWKB} & = & 2m\cosh\left(\vartheta_{n}\right)\label{eq:iwkb_ising}
\end{eqnarray}
The first term in this expansion is given in \cite{FonZam2006} and
it is
\[
S_{1}\left(\vartheta\right)=\frac{1}{\sinh\left(2\vartheta\right)}\left(-\frac{1}{6}\sinh^{2}\left(\vartheta\right)+\frac{5}{24\sinh^{2}\left(\vartheta\right)}+\frac{1}{4\cosh^{2}\left(\vartheta\right)}-\frac{1}{12}\right)
\]
The Bethe-Salpeter equation (with various improvements) leads to a
low energy expansion of the form

\begin{equation}
\frac{\left(m_{n}^{le}\right)^{2}}{4m^{2}}=1+\sum_{k=1}^{\infty}\mu_{k}t^{k}
\end{equation}
with the parameter $t=\lambda^{2/3}$. Different approximations of
the $\mu_{k}$ coefficients taking into account multi-quark corrections
(such as quark mass renormalization and renormalization of the short
range quark-antiquark interaction), and string tension renormalization
can be found in \cite{FonZam2003,Rutkevich2005,FonZam2006,Rutkevich2009}.
The low-energy expansion for the meson mass $\tilde{m}_{n}$ from
the Bethe-Salpeter equation has the form \cite{FonZam2006} 
\begin{eqnarray}
\frac{\tilde{m}_{n}^{2}}{4m^{2}} & = & 1+z_{n}t^{2}+\frac{z_{n}^{2}}{5}t^{4}-\left(\frac{3z_{n}^{3}}{175}+\frac{57}{280}\right)t^{6}+\left(\frac{23z_{n}^{4}}{7875}+\frac{1543z_{n}}{126\text{00}}\right)t^{8}\nonumber \\
 &  & +\frac{13}{1120\pi}t^{9}+\left(-\frac{1894z_{n}^{5}}{3031875}-\frac{23983z_{n}^{2}}{242550}\right)t^{10}+\frac{3313z_{n}}{10080}t^{11}+\dots
\end{eqnarray}
while the radiative corrections modify this expression according to
\begin{equation}
\frac{m_{n}^{le}-\tilde{m}_{n}}{m}=a_{2}t^{2}+\frac{z_{n}}{6}(4c_{2}-a_{2})t^{8}-\frac{B_{2}}{4}t^{9}+O(t^{10})\label{eq:BS_ising}
\end{equation}
where 
\begin{equation}
a_{2}=0.0710809\dots\quad c_{2}=-0.003889\dots
\end{equation}
are the leading order quark mass and string tension renormalization
corrections computed in \cite{FonZam2006} and
\begin{equation}
B_{2}=0.8
\end{equation}
is the leading order interaction renormalization correction obtained
in \cite{Rutkevich2009,Rutkevich2010}.

\subsubsection{Mesonic states in the three state Potts model\label{par:Mesonic-states-in}}

For this latter model, the Bethe-Salpeter has not been carried out
(albeit the setting was established in \cite{RutPottsMes2010}), so
we quote the linear potential quantum mechanics and the WKB results.

\paragraph{h<0}

~

For this sign of the magnetic field there is a single stable vacuum
and two metastable ones. The two-kink configurations are

\[
K_{3\alpha}(\vartheta_{1})K_{\alpha3}(\vartheta_{2})
\]
where $3$ is the stable vacuum, while $\alpha=1,2$ are metastable
vacua. Due to the presence of this degree of freedom allowed by $\alpha$
both antisymmetric and symmetric solutions are allowed. From the simple
quantum mechanical picture of two-kink configurations it is clear
that in the sector of zero total momentum the charge conjugation ($\mathcal{C}$)
parity of kink-antikink and therefore also meson states is equal to
their parity under spatial reflections.

The spectrum predicted by quantum mechanics in the linear potential
is 
\begin{eqnarray}
m_{-,n}^{\left(Ai\right)} & = & m(2+\lambda^{2/3}z_{n})+\mathcal{O}\left(\lambda^{4/3}\right)\\
m_{+,n}^{\left(Ai\right)} & = & m(2+\lambda^{2/3}z'_{n})+\mathcal{O}\left(\lambda^{4/3}\right)
\end{eqnarray}
where $\lambda$ is the dimensionless ratio
\begin{equation}
\lambda=\frac{\beta^{\left(3\right)}|h|}{m^{2}}
\end{equation}
and $-z_{n}$ is the $n$-th zero of the Airy function, while $-z'_{n}$
is the $n$-th zero of its derivative. 

For this case, the WKB quantization is given by \cite{RutPottsMes2010}
\begin{eqnarray}
\frac{\sinh\left(2\vartheta_{n}\right)-2\vartheta_{n}}{\lambda} & = & 2\pi\left(n-\frac{1}{4}\right)+2\arctan\left(\frac{\tanh2\vartheta_{n}}{\sqrt{3}}\right)+i\mathcal{A}(2\vartheta_{n})+O(\lambda)\quad\mathrm{odd}\\
\frac{\sinh\left(2\vartheta_{n}\right)-2\vartheta_{n}}{\lambda} & = & 2\pi\left(n-\frac{3}{4}\right)+2\arctan\left(\frac{\tanh2\vartheta_{n}}{\sqrt{3}}\right)+i\mathcal{A}(2\vartheta_{n})+O(\lambda)\quad\mathrm{even}\\
\mathcal{A}(\vartheta) & = & \log\left(\frac{\sinh(i\pi/3+\vartheta)}{\sinh(i\pi/3-\vartheta)}\right)\\
m_{\pm,n}^{WKB} & = & 2m\cosh\left(\vartheta_{n}\right)
\end{eqnarray}
This includes effects of nontrivial kink-antikink scattering, and
therefore despite being semiclassical it goes beyond the simple quantum
mechanical result above.

\paragraph{h>0}

~

For this sign of the magnetic field there are two stable vacua and
one metastable. The allowed two-kink configurations are
\[
K_{\alpha3}(\vartheta_{1})K_{3\beta}(\vartheta_{2})
\]
Charged meson states correspond to $\alpha\neq\beta$, and their spectrum
is given by the WKB quantization condition \cite{RutPottsMes2010}
is given by
\begin{equation}
\frac{\sinh\left(2\vartheta_{n}\right)-2\vartheta_{n}}{\lambda}=2\pi\left(n-\frac{1}{4}\right)+i\mathcal{A}(2\vartheta_{n})+O(\lambda)
\end{equation}
Note that charged meson single-particle states do not satisfy periodic
boundary conditions on a circle due to $\alpha\neq\beta$, therefore
they cannot be observed in TCSA.

On the other hand, neutral meson states do not exist, as they easily
decompose under the process
\[
K_{\alpha3}(\vartheta_{1})K_{3\alpha}(\vartheta_{2})\rightarrow K_{\alpha\beta}(\vartheta_{2})K_{\beta\alpha}(\vartheta_{1})
\]
where for $\alpha=1,2$ one has $\beta=2,1$, respectively. This process
is allowed by the Chim-Zamolodchikov kink scattering amplitudes \cite{ChimZam1992};
also note that the kinks mediating between the stable vacua $1,2$
are not confined.

\subsubsection{Baryon masses in the three state Potts model\label{par:Baryon-masses-in}}

For the case $h<0$ all the kinks are confined. As a result, one may
have three-kink bound states of the form
\begin{eqnarray*}
 &  & K_{31}(\vartheta_{1})K_{12}(\vartheta_{2})K_{23}(\vartheta_{3})\\
 &  & K_{32}(\vartheta_{1})K_{21}(\vartheta_{2})K_{13}(\vartheta_{3})
\end{eqnarray*}
corresponding to baryons and antibaryons. Both of these particles
have the same spectrum due to charge-conjugation symmetry, and can
be modeled in the form of a quantum mechanical three-body system.
The low energy estimates for the baryon masses were recently obtained
by Rutkevich \cite{Rut2015} with the result
\begin{equation}
M_{n}^{\pm}=m(3+\left(\beta^{\left(3\right)}|h|/m^{2}\right)^{2/3}\epsilon_{n}^{\pm})+\mathcal{O}\left(|h|^{4/3}\right)\label{eq:rutkevich_predictions}
\end{equation}
where the $\pm$ correspond to parity under space reflection, with
the following numerical values of $\epsilon$ for the first three
states: 
\begin{equation}
\begin{array}{ccc}
\epsilon_{1}^{+}=4.602 & \epsilon_{2}^{+}=5.912 & \epsilon_{3}^{+}=7.098\\
\epsilon_{1}^{-}=6.650 & \epsilon_{2}^{-}=7.734 & \epsilon_{3}^{-}=8.753
\end{array}\label{eq:rutkevich_epsilons}
\end{equation}
There are no baryons for $h>0$ as the kinks between the two stable
vacua are not confined.

\section{TCSA, RG and extrapolation\label{sec:TCSA:-RG-and}}

\subsection{TCSA for the Ising and Potts field theories}

\subsubsection{Scaling Ising model}

The Hilbert space of any conformal field theory can be decomposed
into products of irreducible representations of the left and right
moving Virasoro algebras, which can be specified by giving their left
and right conformal weights as 
\begin{equation}
\mathcal{S}_{h,\bar{h}}=\mathcal{V}_{h}\otimes\mathcal{V}_{\bar{h}}
\end{equation}
and every such sector corresponds to a primary field $\Phi_{h,\bar{h}}$
. For the Ising model with central charge $c=1/2$ the full Hilbert
space is 
\begin{eqnarray}
\mathcal{H}^{(2)} & = & \mathcal{S}_{0,0}\oplus\mathcal{S}_{\frac{1}{2},\frac{1}{2}}\\
 &  & \oplus\mathcal{S}_{\frac{1}{16},\frac{1}{16}}
\end{eqnarray}
where the sectors on the first line are even, the ones on the second
line are odd. The Hamiltonian is 
\begin{equation}
H=H_{CFT}^{\left(2\right)}+\tau\int dx\varepsilon+h\int dx\sigma
\end{equation}
where 
\begin{equation}
\varepsilon=\Phi_{\frac{1}{2},\frac{1}{2}}\quad\sigma=\Phi_{\frac{1}{16},\frac{1}{16}}
\end{equation}
Note that for this model the values $h$ and $-h$ are physically
equivalent since they are related by the $\mathbb{Z}_{2}$ symmetry
of the conformal field theory.

For the Ising model we used the following level cut-offs with the
dimensions of the truncated Hilbert space indicated below:

\begin{center}
\begin{tabular}{|c|c|c|c|c|c|c|c|c|c|c|}
\hline 
$n$ & $6$ & $7$ & $8$ & $9$ & $10$ & $11$ & $12$ & $13$ & $14$ & $15$\tabularnewline
\hline 
\hline 
$\dim$ & $77$ & $127$ & $213$ & $338$ & $551$ & $840$ & $1330$ & $1994$ & $3023$ & $4476$\tabularnewline
\hline 
\end{tabular}
\par\end{center}

\subsubsection{Scaling 3-state Potts model}

The scaling limit of Potts model at the critical point is a minimal
conformal field theory with central charge 
\begin{equation}
c=\frac{4}{5}
\end{equation}
\cite{Belavin:1984vu,Dotsenko:1984if}. The Hilbert space of the Potts
model is the $D_{4}$ modular invariant \cite{Cappelli:1986hf} 
\begin{eqnarray}
\mathcal{H} & = & \mathcal{H}_{0}\oplus\mathcal{H}_{+}\oplus\mathcal{H}_{-}\oplus\mathcal{H}_{1}\label{eq:Dinvariantsectors}
\end{eqnarray}
where 
\begin{eqnarray}
\mathcal{H}_{0} & = & \mathcal{S}_{0,0}\oplus\mathcal{S}_{\frac{2}{5},\frac{2}{5}}\oplus\mathcal{S}_{\frac{7}{5},\frac{7}{5}}\oplus\mathcal{S}_{3,3}\nonumber \\
\mathcal{H}_{\pm} & = & \mathcal{S}_{\frac{1}{15},\frac{1}{15}}^{\pm}\oplus\mathcal{S}_{\frac{2}{3},\frac{2}{3}}^{\pm}\nonumber \\
\mathcal{H}_{1} & = & \mathcal{S}_{\frac{2}{5},\frac{7}{5}}\oplus\mathcal{S}_{\frac{7}{5},\frac{2}{5}}\oplus\mathcal{S}_{0,3}\oplus\mathcal{S}_{3,0}\label{eq:Hilbert_space_sectors}
\end{eqnarray}
The $D_{4}$ conformal field theory is invariant under the permutation
group $\mathbb{S}_{3}$ generated by two elements $\mathcal{Z}$ and
$\mathcal{C}$ with the relations
\begin{equation}
\mathcal{Z}^{3}=1\qquad\mathcal{C}^{2}=1\qquad\mathcal{CZC}=\mathcal{Z}^{-1}
\end{equation}
which have the signatures
\begin{equation}
\mbox{sign }\mathcal{Z}=+1\qquad\mbox{sign }\mathcal{C}=-1
\end{equation}
The sectors in $\mathcal{H}_{0}$ of (\ref{eq:Dinvariantsectors})
are invariant under the action of the permutation group $\mathbb{S}_{3}$,
the ones in $\mathcal{H}_{\pm}$ form the two-dimensional irreducible
representation, which is characterized by the following action of
the generators:
\begin{eqnarray}
\mathcal{C}|\pm\rangle & = & \pm|\mp\rangle\nonumber \\
\mathcal{Z}|\pm\rangle & = & \cos\left(\frac{2\pi}{3}\right)|\pm\rangle\pm\sin\left(\frac{2\pi}{3}\right)|\mp\rangle
\end{eqnarray}
while those in $\mathcal{H}_{1}$ transform according to the signature
representation of $\mathbb{S}_{3}$. 

The Hamiltonian is given by

\begin{equation}
H=H_{CFT}^{\left(3\right)}+\tau\int dx\varepsilon+h\int dx\sigma
\end{equation}
where 
\begin{equation}
\varepsilon=\Phi_{\frac{2}{5},\frac{2}{5}}\quad\sigma=\Phi_{\frac{1}{15},\frac{1}{15}}^{+}
\end{equation}
In our considerations an important role will be played by charge conjugation
parity: the even sector under $\mathcal{C}$ consists of $\mathcal{H}_{0}$
and $\mathcal{H}_{+}$, while the odd sector consists of $\mathcal{H}_{1}$
and $\mathcal{H}_{-}$, with the dimensions as a function of the level
cut-off $n$ given below:

\begin{center}
\begin{tabular}{|c|c|c|c|c|c|c|}
\hline 
$n$ & $6$ & $7$ & $8$ & $9$ & $10$ & $11$\tabularnewline
\hline 
\hline 
dim, even sector & $634$ & $1210$ & $2426$ & $4437$ & $8258$ & $14545$\tabularnewline
\hline 
dim, odd sector & $816$ & $1572$ & $3039$ & $5592$ & $10121$ & $\left(17904\right)$\tabularnewline
\hline 
\end{tabular}
\par\end{center}

In the case of the odd sector the left/right descendent levels must
be different for some of the fields the right in order to get spinless
fields. In our convention, the truncation level is chosen to agree
with the smaller of the descendent levels. In the extrapolations we
used level cut-offs from $6$ to $11$ in the even sector and from
$6$ to $10$ in the odd sector.

\subsection{TCSA numerics: conventions}

The conformal Hamiltonian on a finite circle of circumference $R$
can be written as
\begin{equation}
H_{CFT}^{\left(q\right)}=\frac{2\pi}{R}\left(L_{0}+\bar{L}_{0}-\frac{c}{12}\right)
\end{equation}
In our TCSA calculations we consider the zero momentum sectors for
which $L_{0}-\bar{L}_{0}=0$ , and impose a level cutoff on the spectrum
of states according to
\begin{equation}
L_{0}\leq n
\end{equation}
with $n$ a positive integer. 

The computations are all performed in units of the kink mass $m$
of the $h=0$ model, which means the volume is measured in dimensionless
units $r=mR$ and the dimensionless energy levels are $e=E/m$. The
finite volume energy levels are given as functions $e_{i}(r)$, with
$i$ indexing the different levels; by convention the vacuum is taken
to correspond to $i=0$. Since we are interested in the mass spectrum,
the relevant quantities are the relative energy levels
\begin{equation}
\tilde{e}_{i}(r)=e_{i}(r)-e_{0}(r)
\end{equation}
One can also introduce the scaling function defined by
\begin{equation}
d_{i}(r)=\frac{r}{2\pi}e_{i}(r)
\end{equation}
In the conformal field theory, the scaling function is a constant
and is given by the eigenvalue of the operator
\begin{equation}
L_{0}+\bar{L}_{0}-\frac{c}{12}
\end{equation}
corresponding to the given level. In the off-critical model the leading
term of $e_{i}(r)$ for large $r$ is given by the bulk energy and
is therefore generally linear (the Ising case is an exception with
a logarithmic contribution), and so the scaling functions $d_{i}(r)$
grow as $r^{2}$. On the other hand, the bulk contribution is universal
for all levels, and in the absence of certain ultraviolet divergences
(which is the case for the models considered here, as none of the
perturbing operators has conformal weight $h\geq3/4$ \cite{GiokasWatts2011}),
the relative energy levels $\tilde{e}_{i}(r)$ go to a constant for
large $r$.

\subsection{Renormalization group improvement}

To leading order in the level cutoff $n$, the cutoff dependence of
the TCSA can be canceled by allowing the couplings to run according
to renormalization group equations derived from second order perturbation
theory. For a Hamiltonian of the form 
\begin{equation}
H=\frac{2\pi}{R}\left(L_{0}+\bar{L}_{0}-\frac{c}{12}\right)+\sum_{a}\lambda_{a}\int_{0}^{R}dx\Phi_{a}(x)
\end{equation}
the leading RG equations are \cite{GiokasWatts2011,Rychkov2015a}
\begin{eqnarray}
\tilde{\lambda}_{c}(n)-\tilde{\lambda}_{c}(n-1) & = & \sum_{a,b}\tilde{\lambda}_{a}(n)\tilde{\lambda_{b}}(n)C_{ab}^{c}\frac{n^{2h_{abc}-3}}{2\Gamma(h_{abc})^{2}}\left(1+O(1/n)\right)\\
 &  & \tilde{\lambda}_{a}=\frac{\lambda_{a}R^{2-2h_{a}}}{(2\pi)^{1-2h_{a}}}\nonumber \\
 &  & h_{abc}=h_{a}+h_{b}-h_{c}\nonumber 
\end{eqnarray}
where $C_{ab}^{c}$ are the CFT operator product expansion coefficients:
\begin{equation}
\Phi_{a}(z,\bar{z})\Phi_{b}(0,0)=\sum_{c}\frac{C_{ab}^{c}\Phi_{c}(0,0)}{z^{h_{a}+h_{b}-h_{c}}\bar{z}^{\bar{h}_{a}+\bar{h}_{b}-\bar{h}_{c}}}
\end{equation}
In the perturbation theory calculation, however, large denominators
may appear due to the $r^{2}$ dependence of the scaling functions.
This can be compensated for by taking into account the universal part
of all scaling functions, by modifying the RG equations following
the prescription in \cite{Rychkov2015a,Rychkov2015b} 
\begin{equation}
\tilde{\lambda}_{c}(n)-\tilde{\lambda}_{c}(n-1)=\frac{1}{2n-d_{0}(r)}\sum_{a,b}\tilde{\lambda}_{a}(n)\tilde{\lambda_{b}}(n)C_{ab}^{c}\frac{n^{2h_{abc}-2}}{\Gamma(h_{abc})^{2}}\left(1+O(1/n)\right)\label{eq:RG_eqs}
\end{equation}
where the vacuum scaling function $d_{0}(r)$ can be estimated by
its TCSA value at the starting cutoff for the RG run. In a unitary
field theory it can be argued that the vacuum scaling function is
always negative, so this does not introduce any new singularities.
In essence this modification means regrouping some potentially large
$1/n$ corrections into the RG flow. 

Note that this prescription also gives a running coupling for the
identity, which leads to an additive renormalization constant for
all energy levels. Since we consider only relative energy levels,
this contribution can be omitted.

\subsection{Extrapolation}

The higher $1/n$ terms give state dependent corrections corresponding
to non-local counter terms \cite{GiokasWatts2011,Rychkov2015a,Rychkov2015b}.
Their construction is quite involved and they are not known in a fully
analytic form yet. However, there is an efficient shortcut that is
sufficient for the purposes of the present work. The leading terms
already incorporated in the RG equations yield cut-off corrections
after integration with the scaling form
\begin{equation}
n^{2h_{abc}-2}
\end{equation}
corresponding to the occurrence of $\Phi_{c}$ in the OPE $\Phi_{a}\Phi_{b}$. 

In the Ising model, the leading exponents can be summarized as:

\begin{center}
\begin{tabular}{|c|c|c|c|}
\hline 
$\Phi_{a}\Phi_{b}\backslash\Phi_{c}$ & $\mathbf{1}$ & $\sigma$ & $\epsilon$\tabularnewline
\hline 
\hline 
$\sigma\sigma$ & $-\frac{7}{4}$ & -- & $-\frac{11}{4}$\tabularnewline
\hline 
$\sigma\epsilon$ & -- & $-\frac{15}{8}$ & --\tabularnewline
\hline 
$\epsilon\epsilon$ & $0$ & -- & --\tabularnewline
\hline 
\end{tabular}
\par\end{center}

The exponent $0$ corresponds to a logarithmic divergence in the ground
state energy, which cancels from the relative energy levels. The exponent
$-7/4$ also corresponds to ground state renormalization and therefore
also cancels. The two other exponents are taken care of by the running
couplings. Therefore it is only necessary to take into account the
highest subleading $1/n$ corrections, which lead to a residual cut-off
dependence of the form
\begin{equation}
\tilde{e}_{i}^{(n)}(r)=\tilde{e}_{i}(r)+\frac{A_{i}(r)}{n}+\frac{B_{i}(r)}{n^{2}}+O(n^{-11/4})\label{eq:ising_residual}
\end{equation}
where both subleading terms come from $1/n$ corrections to the $\epsilon\epsilon\mathbf{1}$
term. 

In the Potts model, the leading exponents are

\begin{center}
\begin{tabular}{|c|c|c|c|}
\hline 
$\Phi_{a}\Phi_{b}\backslash\Phi_{c}$ & $\mathbf{1}$ & $\sigma$ & $\epsilon$\tabularnewline
\hline 
\hline 
$\sigma\sigma$ & $-\frac{26}{15}$ & $-\frac{28}{15}$ & $-\frac{38}{15}$\tabularnewline
\hline 
$\sigma\epsilon$ & -- & $-\frac{6}{5}$ & --\tabularnewline
\hline 
$\epsilon\epsilon$ & $-\frac{2}{5}$ & -- & --\tabularnewline
\hline 
\end{tabular}
\par\end{center}

\begin{figure}
\begin{centering}
\subfloat[First excited state]{\centering{}\includegraphics[scale=0.25]{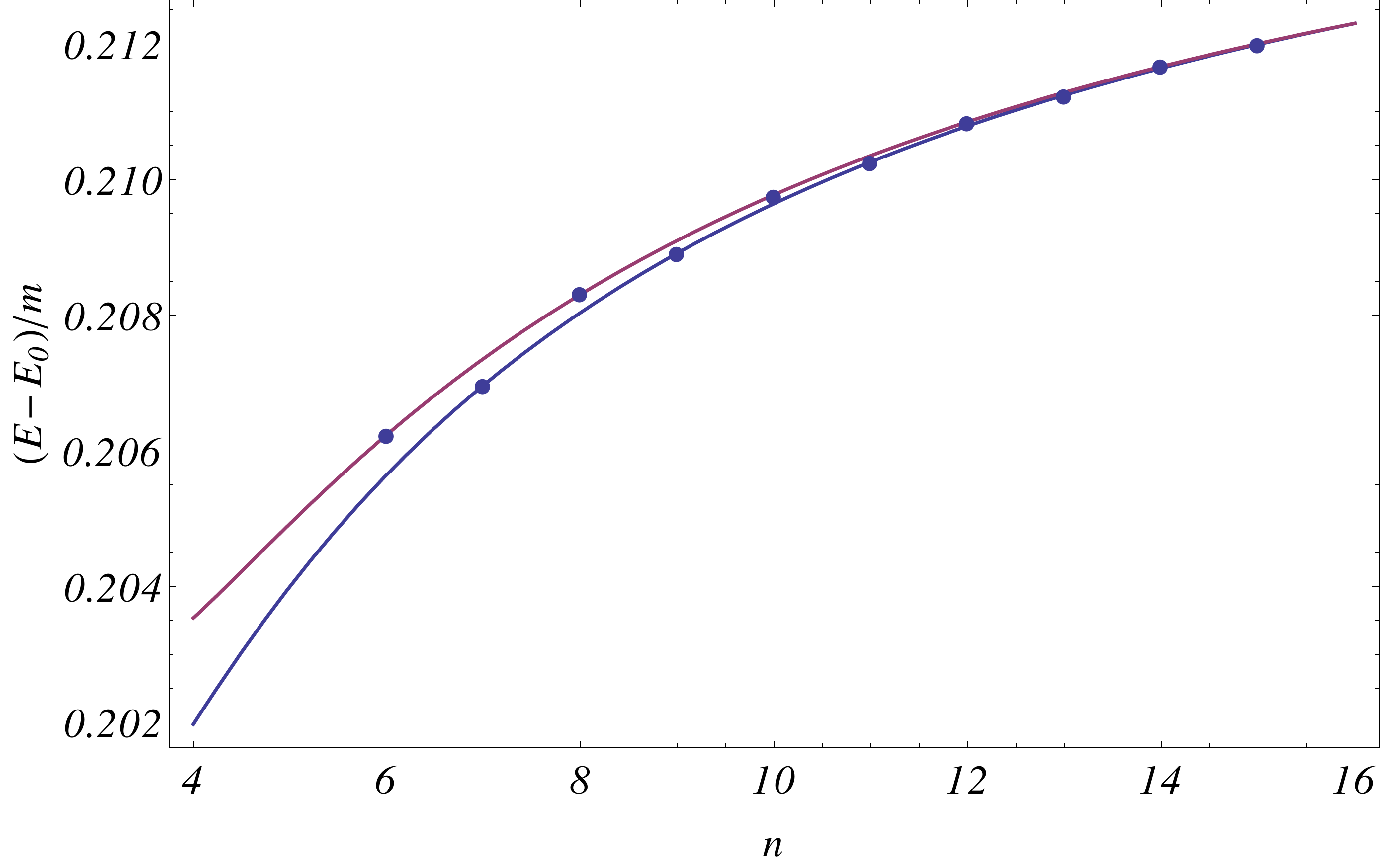}}~~\subfloat[Second excited state]{\centering{}\includegraphics[scale=0.25]{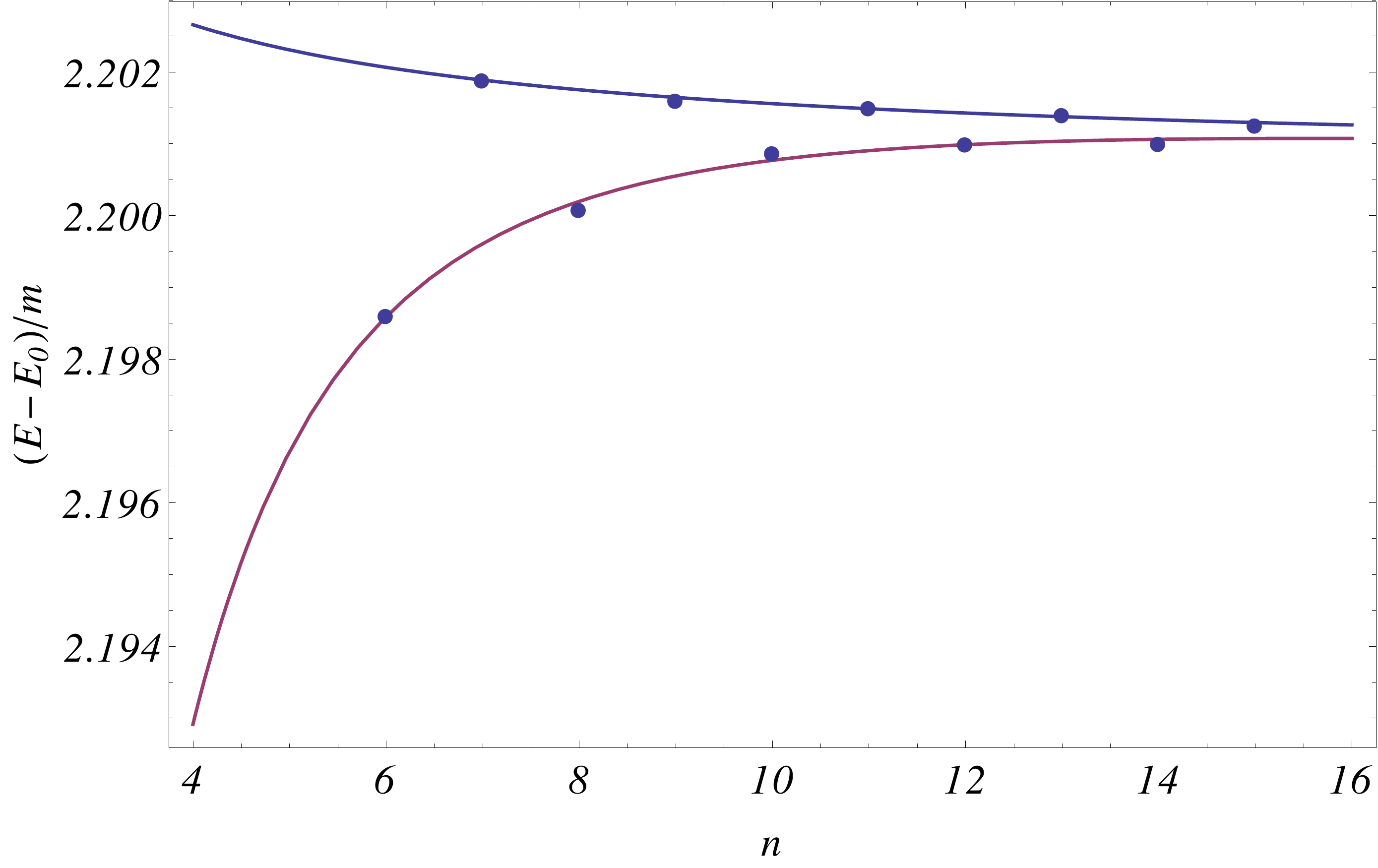}}
\par\end{centering}

\begin{centering}
\subfloat[Third excited state]{\centering{}\includegraphics[scale=0.25]{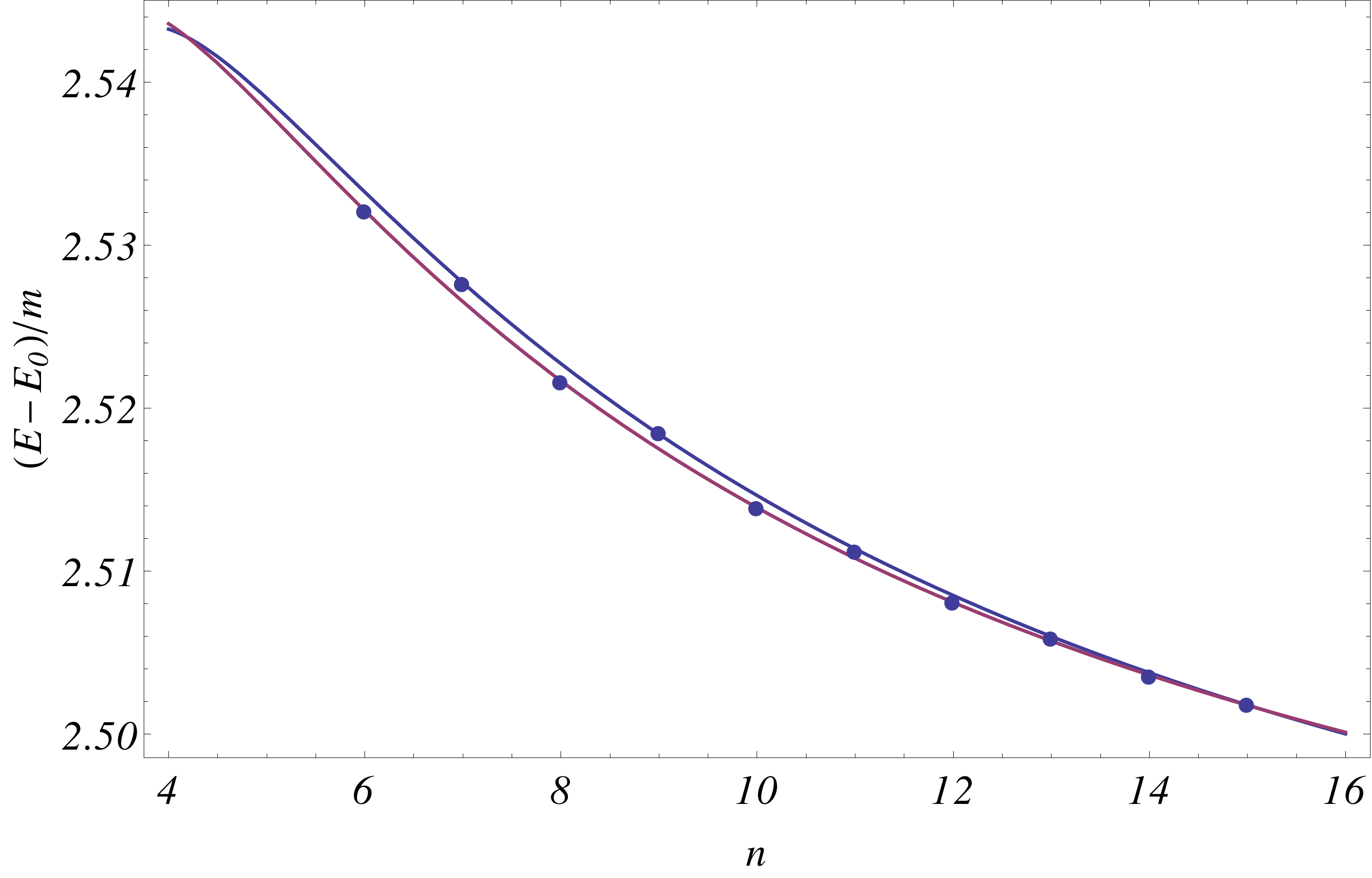}}~~\subfloat[Forth excited state]{\centering{}\includegraphics[scale=0.25]{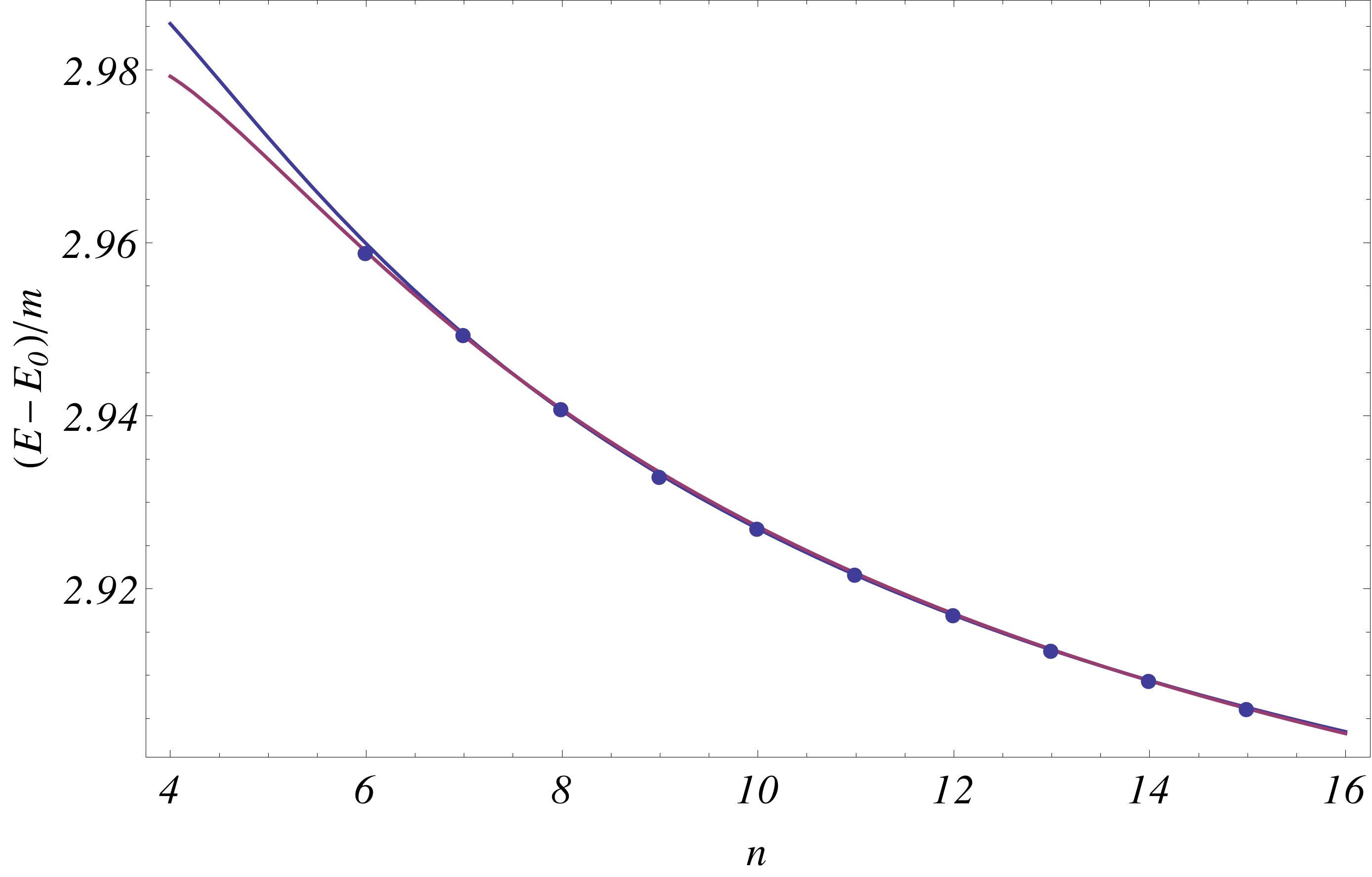}}
\par\end{centering}

\protect\caption{Extrapolation fits of the relative energy levels in the Ising model
in the ferromagnetic phase with magnetic field $\tilde{h}=0.008$
for the first four excited states at dimensionless volume $mR=10$.\label{fig:Energy-against-the}}
\end{figure}

\begin{figure}
\begin{centering}
\subfloat[First excited state]{\centering{}\includegraphics[scale=0.25]{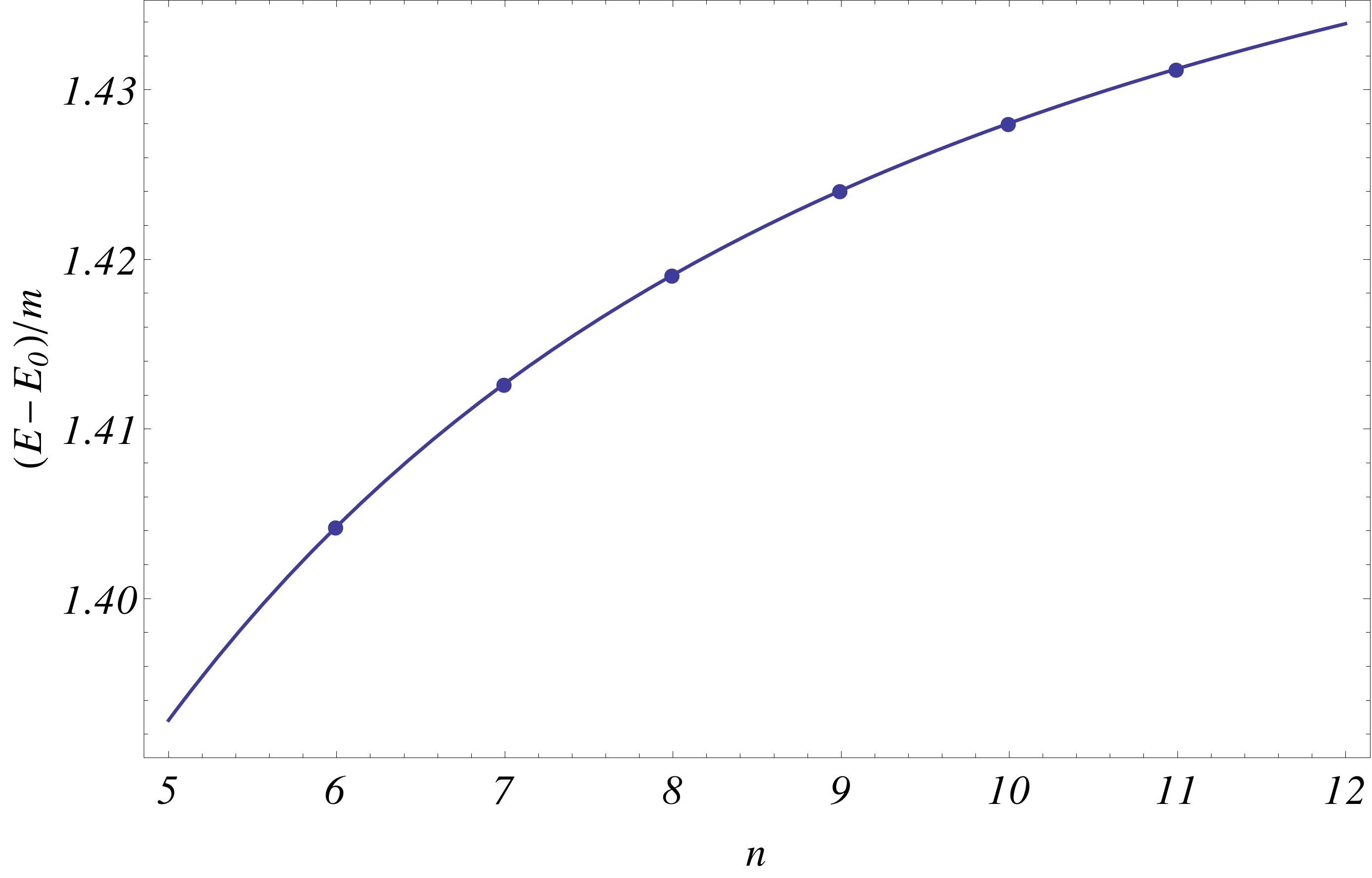}}~~\subfloat[Second excited state]{\centering{}\includegraphics[scale=0.25]{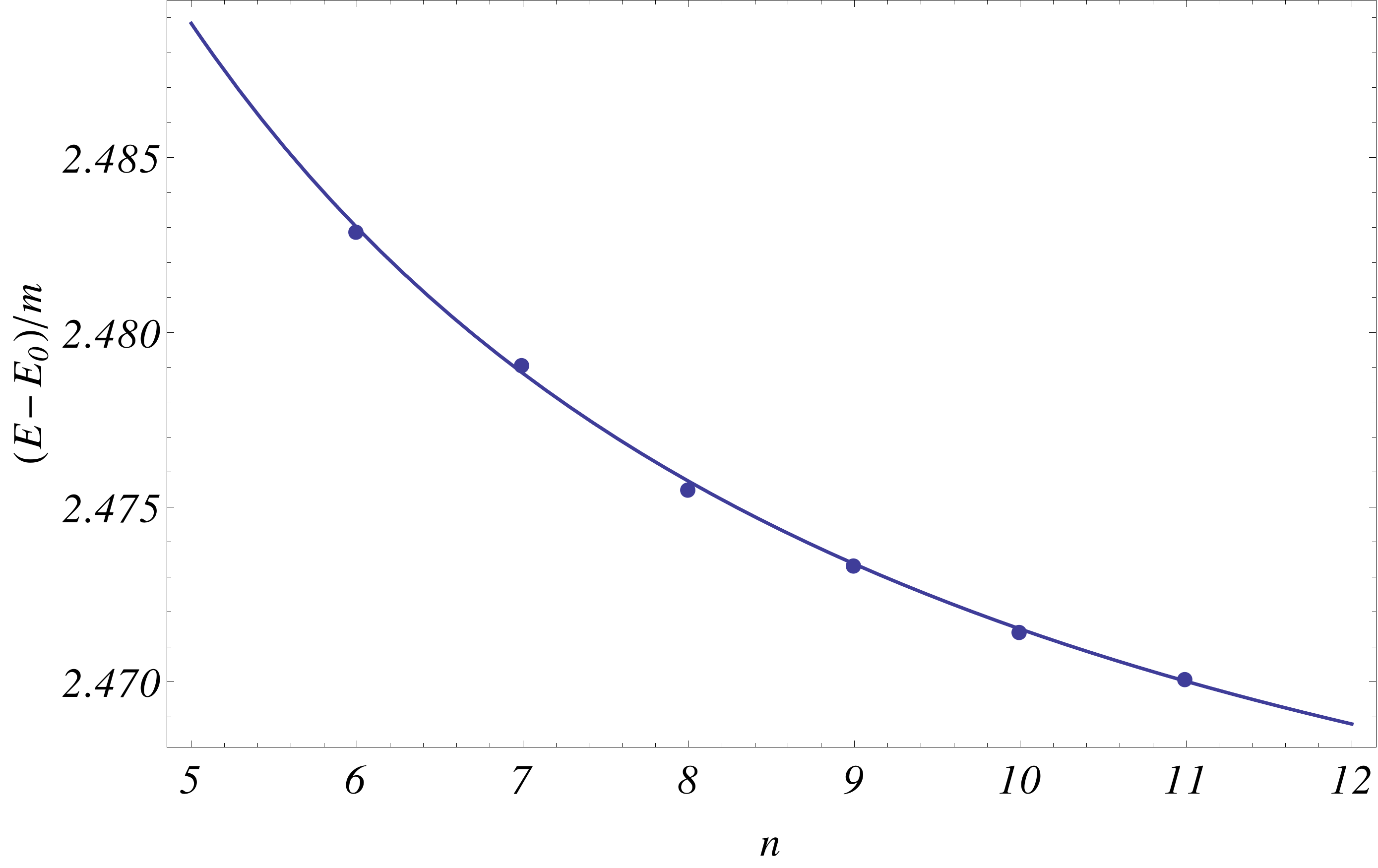}}
\par\end{centering}

\begin{centering}
\subfloat[Third excited state]{\centering{}\includegraphics[scale=0.25]{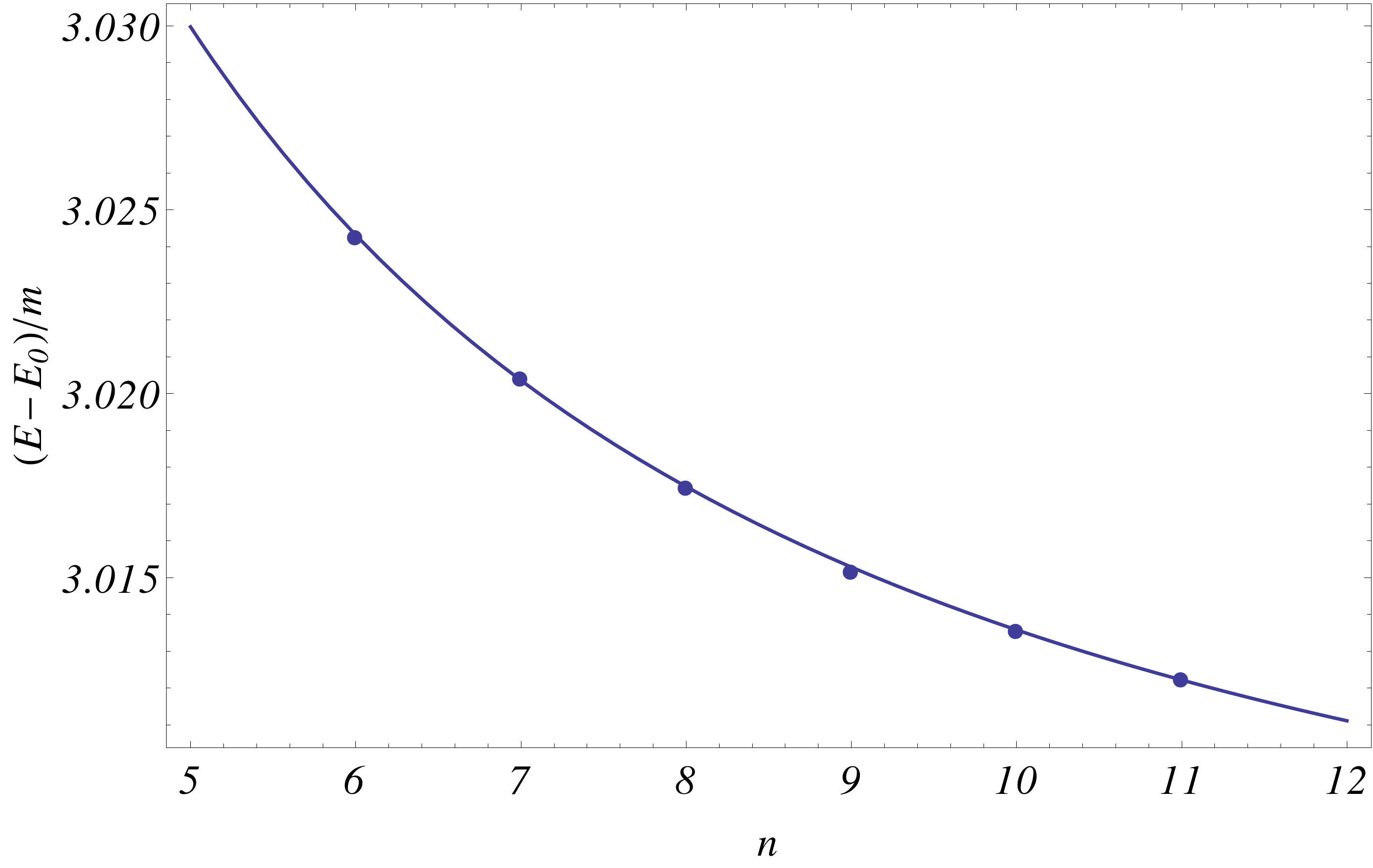}}~~\subfloat[Forth excited state]{\centering{}\includegraphics[scale=0.25]{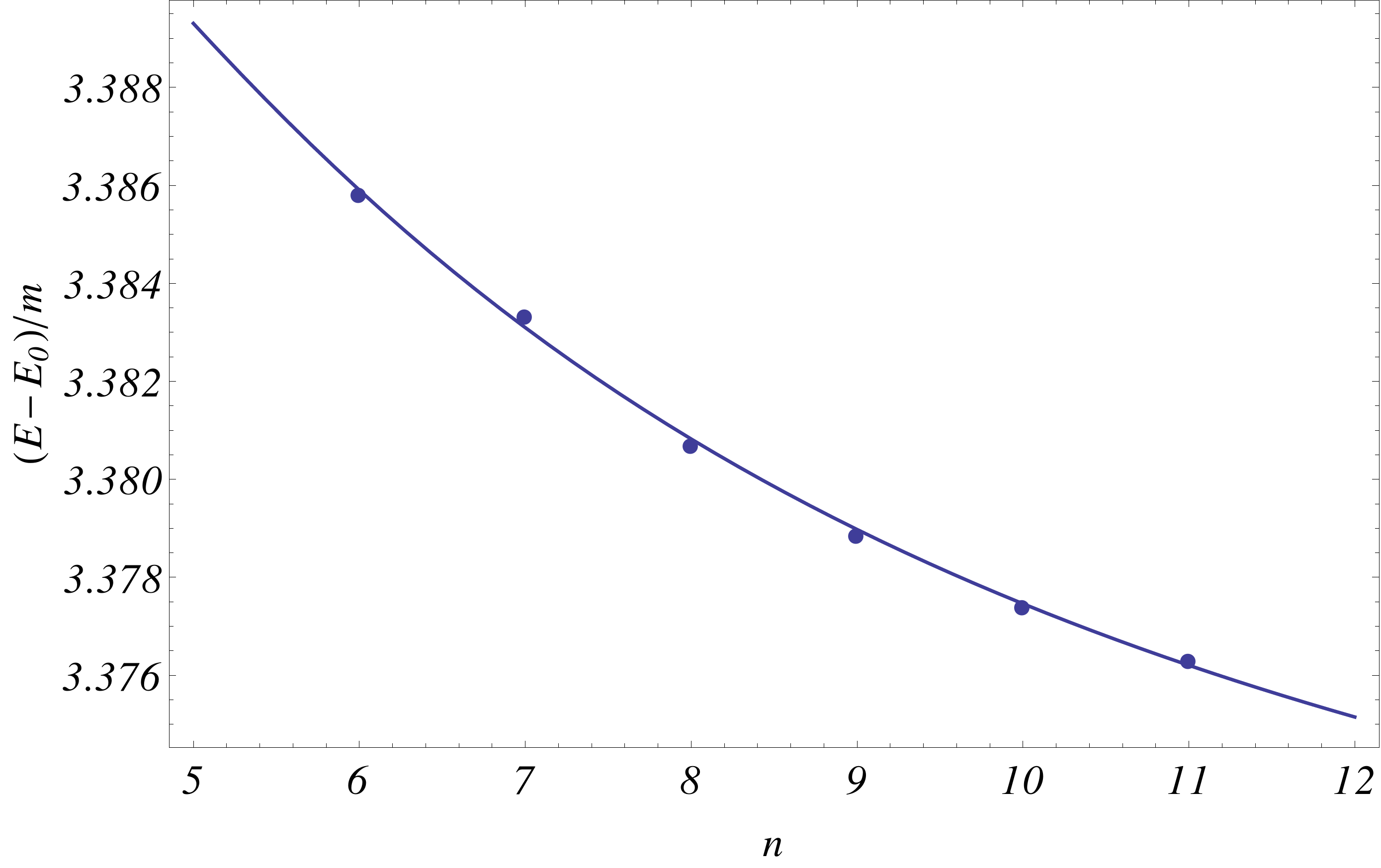}}
\par\end{centering}

\protect\caption{Extrapolation fits of the relative energy levels in the Potts model
in the ferromagnetic phase with magnetic field $\tilde{h}=-0.05$
for the first four excited states at dimensionless volume $mR=10$.\label{fig:Extrapol_Potts}}
\end{figure}

\noindent (where some fields contained in the OPE which give even
higher exponents have been omitted) and the residual cut-off dependence
is
\begin{equation}
\tilde{e}_{i}^{(n)}(r)=\tilde{e}_{i}(r)+\frac{A_{i}(r)}{n^{7/5}}+\frac{B_{i}(r)}{n^{11/5}}+O(n^{-12/5})\label{eq:potts_residual}
\end{equation}
Our prescription for the RG-TCSA is as follows. In units of $m$,
$\tau$ is just a fixed dimensionless number given by $\kappa^{(q)}$.
Therefore the TCSA has two dimensionless parameters, one of which
is given by value of $h$ in units of $m$, i.e. the ratio 
\begin{equation}
\tilde{h}=h/m^{2-2h_{\sigma}}
\end{equation}
and the dimensionless volume parameter $r=mR$. For any value of $r$
and $m$ the physical values of the perturbed CFT couplings are 
\begin{eqnarray}
\tilde{\lambda}_{\epsilon} & = & -\frac{\kappa^{(q)}r^{2-2h_{\epsilon}}}{(2\pi)^{1-2h_{\epsilon}}}\\
\tilde{\lambda}_{\sigma} & = & \frac{h}{m^{2-2h_{\sigma}}}\frac{r^{2-2h_{\sigma}}}{(2\pi)^{1-2h_{\sigma}}}\nonumber 
\end{eqnarray}
Taking these as initial conditions at $n=\infty$, the couplings can
be run according to the RG equations (\ref{eq:RG_eqs}) to determine
their value at the given cut-off $n$. In practice one can approximate
the difference equations by substituting
\begin{equation}
\tilde{\lambda}_{c}(n)-\tilde{\lambda}_{c}(n-1)\rightarrow\frac{d\tilde{\lambda}_{c}}{dn}
\end{equation}
and solving the resulting differential equations numerically. We remark
that in all our calculations the couplings ran very little so this
has practically no effect, but as a matter of principle it must be
done before we proceed to extrapolation. Once the RG eliminated all
the leading cut-off dependencies, the renormalized TCSA Hamiltonian
can be numerically diagonalized and then the residual cut-off dependence
eliminated by fitting (\ref{eq:ising_residual}) for the Ising and
(\ref{eq:potts_residual}) for the Potts case. In the case of the
Ising model it turns out that the residual cut-off dependence alternates
in sign between odd and even cut-offs, so the data for even and odd
values of $n$ were fitted separately, as demonstrated in figure \ref{fig:Energy-against-the}.
For the case of the Potts model no such alternation was observed,
and the data could be reliably extrapolated including both even and
odd values of the level cut-off $n$ as illustrated in \ref{fig:Extrapol_Potts}.
We also remark that in the Potts case we took into account the two
exponents indicated in (\ref{eq:potts_residual}); the $11/5$ and
$12/5$ exponents are too close together, and their effect is to small
compared to the leading $7/5$ to include both in the fit. 

\section{Results\label{sec:Results}}

\subsection{Testing ground: Ising model with magnetic field}

\subsubsection{False vacuum}

\begin{figure}
\begin{centering}
\subfloat[$\tilde{h}=0.008$]{\centering{}\includegraphics[scale=0.35]{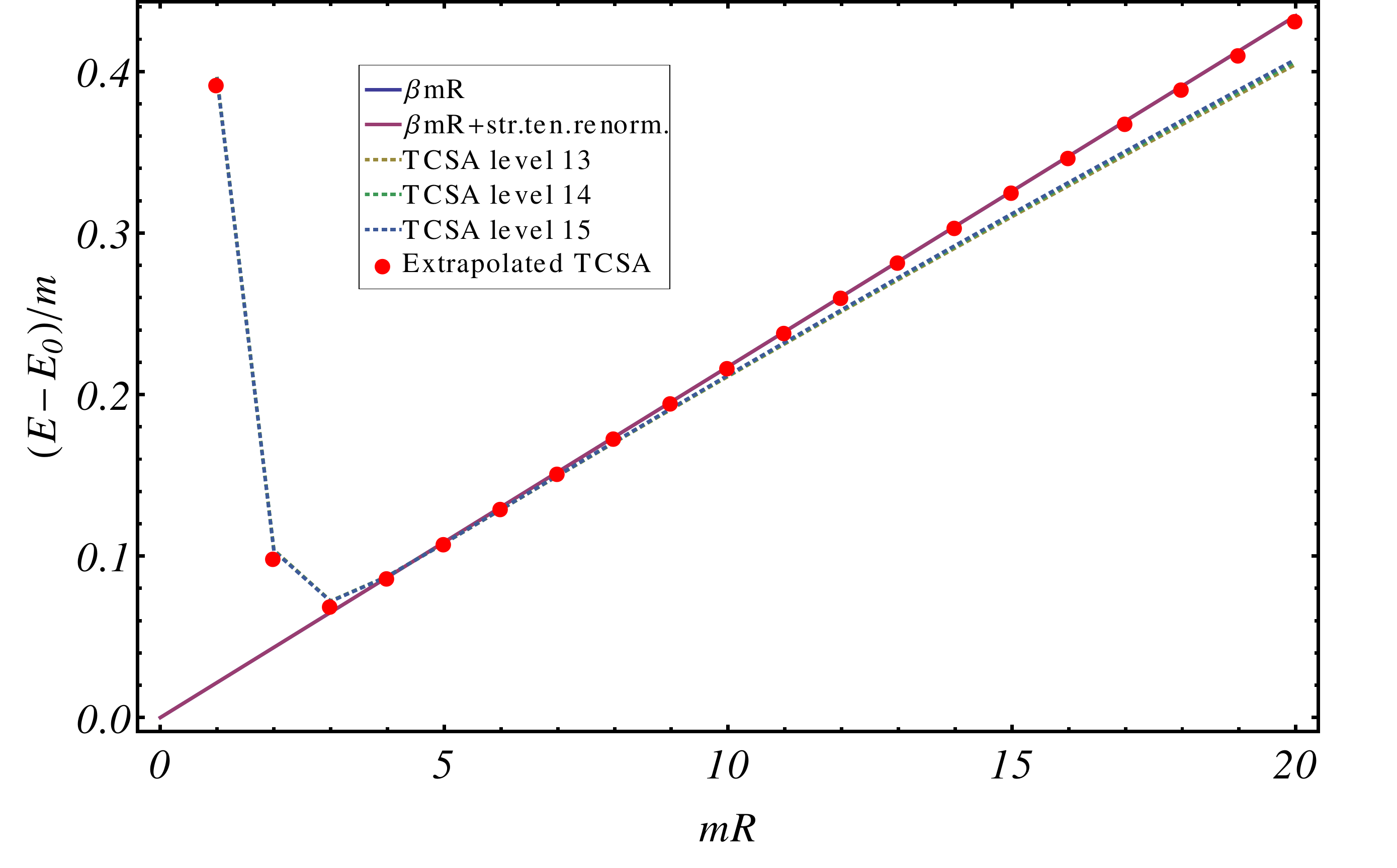}}\\
\subfloat[$\tilde{h}=0.03$]{\centering{}\includegraphics[scale=0.35]{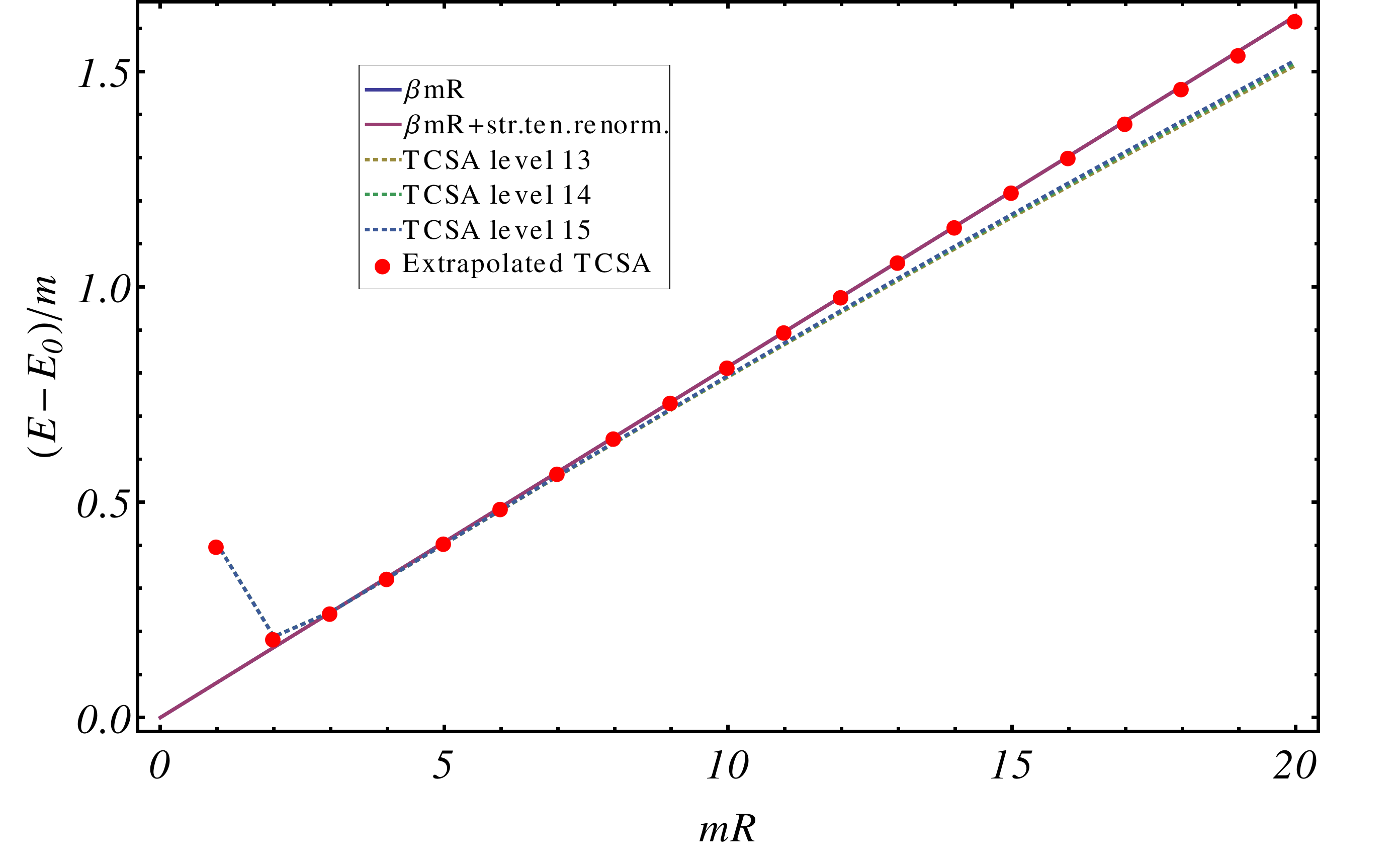}}\\

\par\end{centering}

\centering{}\subfloat[$\tilde{h}=0.05$]{\centering{}\includegraphics[scale=0.35]{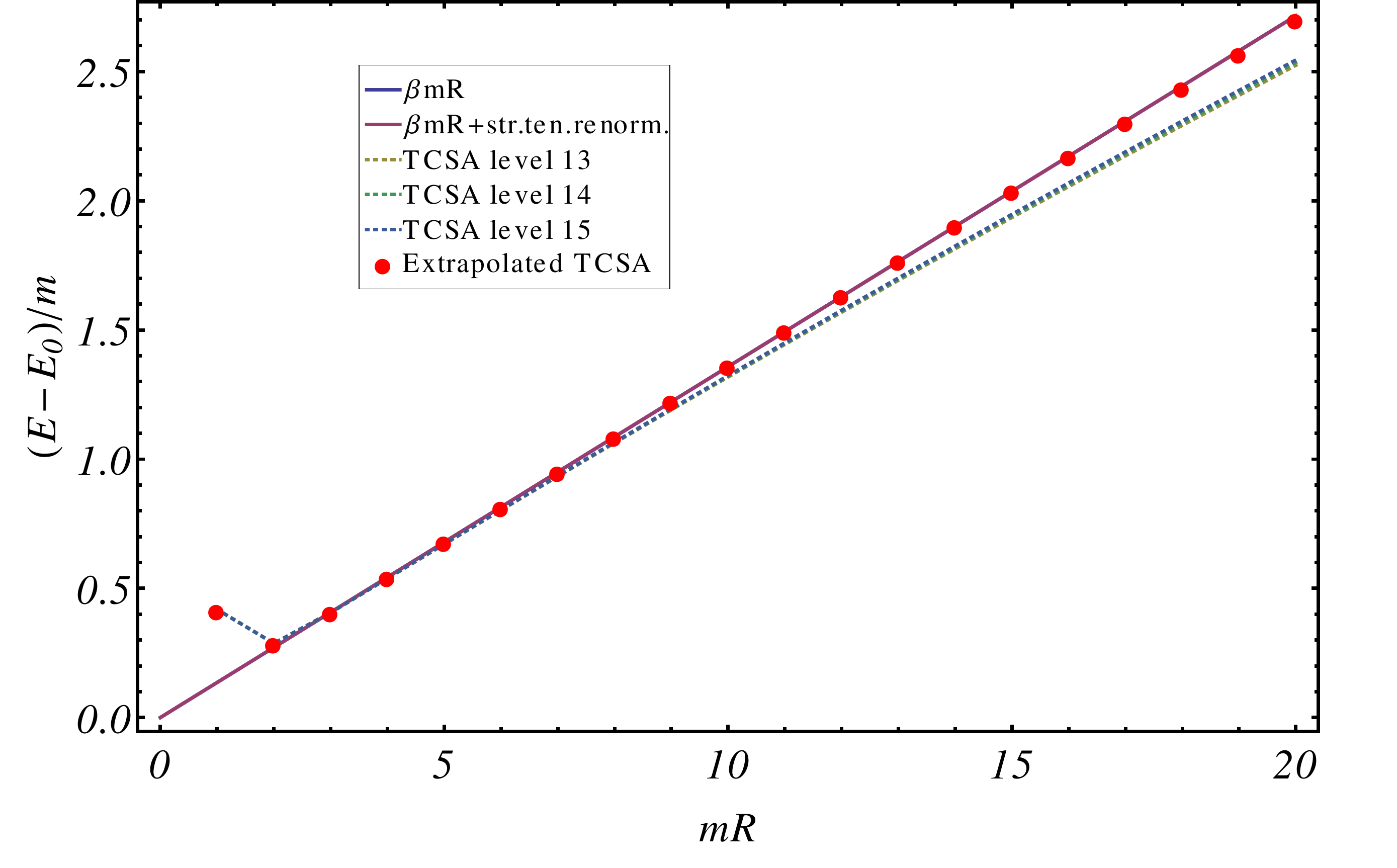}}\protect\caption{\label{fig:Energy-of-the}Energy of the false vacuum for different
values of the magnetic field in the Ising model. Lines are the theoretical
predictions: blue is for (\ref{eq:fv_endensity}), while purple one
takes into account the string tension renormalization (the two lines
are almost indistinguishable in the graphs). Dotted lines for different
cut-off data, red dots are extrapolated (average of even and odd cut-off
extrapolations)}
\end{figure}

From the TCSA data it is possible to evaluate the energy density of
the false vacuum relative to the stable one. For smaller volume, the
TCSA converges fast, but for greater volumes the efficiency of the
extrapolation procedure is apparent. The theoretical predictions can
be calculated using (\ref{eq:fv_endensity}) where the renormalized
string tension given in \cite{FonZam2006} has also been taken into
account. However, at the present precision the two predictions cannot
be distinguished. Our results are illustrated in Figure \ref{fig:Energy-of-the}.

Note that the false vacuum level is not a continuous level in the
volume, therefore its linear rise does not in fact contradict the
statement that all relative energy levels $\tilde{e}_{i}(r)$ go to
a constant for $r\rightarrow\infty$. The metastable states are seen
as level avoidances in finite volume \cite{Luscher:1991cf,Pozsgay:2006wb};
this is best demonstrated by the false meson level in the Potts model,
shown in subsection \ref{sub:Meson-masses}. For the false vacuum,
however, the level avoidances are confined to small enough regions,
so that the level can be glued together from the pieces. In addition,
for small coupling the first level avoidance is at a much larger volume
than shown in the plots. 

\subsubsection{Meson masses}

The meson masses can be estimated by first extrapolating the levels
separately for even and odd cut-offs, and taking the average of the
two results. For a more precise measurement the numerical procedure
was slightly modified by keeping only the $n=10\dots15$ data and
using only the $1/n$ term in (\ref{eq:ising_residual}). The reason
for this is that the meson level data do not allow fitting the $1/n^{2}$
term with a sufficient precision.

Selecting a given extrapolated one-particle level, one then finds
the volume where the level is the most flat. The data still contain
exponential finite size effects, which can be suppressed by fitting
the part of the extrapolated $n$th meson level just before the flat
portion by an exponential function 
\begin{equation}
\tilde{e}_{n}(r)=m_{n}+A_{n}e^{-B_{n}r}\label{eq:expfit}
\end{equation}
and taking $m_{n}$ as the estimated mass of the $n$th meson. The
results are shown in Table \ref{tab:Meson-masses-ising} (we note
that the data did not permit the volume extrapolation for $\tilde{h}=0.4$,
so the TCSA numbers quoted there are just the value of the level at
the point where it is most flat). The theoretical prediction ``Airy''
is given by eqn. (\ref{eq:Airy_ising}), while BS and WKB given by
(\ref{eq:wkb_ising}) and (\ref{eq:BS_ising}), respectively; the
last line (iWKB) corresponds to taking into account the first correction
in (\ref{eq:iwkb_ising}). One can see that the prediction from simple
quantum mechanics in a linear potential (``Airy'') is generally
only precise to a percent level or even worse for large magnetic fields,
while WKB is an order of magnitude better. The theoretically rather
involved Bethe-Salpeter approach only improves on WKB for very low
mass mesons, while for higher masses the WKB is generally better.
It is also clear that for $\tilde{h}\gtrsim0.2$ the only working
theoretical framework is the WKB method. 

The upshot is that for all practical purposes WKB can be taken as
the most reliable description of the spectrum over all the parameter
range: it gives an approximation within $10^{-3}$ relative precision.
This is an important lesson given that there are no Bethe-Salpeter
predictions available in the three-state Potts case yet; however,
we can take the WKB as an accurate prediction for comparison with
the three-state Potts meson data.

\begin{table}
\begin{centering}
\begin{tabular}{|c||l|c|c|c|}
\hline 
$\tilde{h}$ &  & $m_{1}$ & $m_{2}$ & $m_{3}$\tabularnewline
\hline 
\hline 
\multirow{5}{*}{$0.0175$} & TCSA & $2.303$ & $2.526$ & $2.71$\tabularnewline
\cline{2-5} 
 & Airy & $2.3068$ & $2.5364$ & $2.7243$\tabularnewline
\cline{2-5} 
 & BS & $2.3021$ & $2.5228$ & $2.7005$\tabularnewline
\cline{2-5} 
 & WKB & $2.3000$ & $2.5223$ & $2.7003$\tabularnewline
\cline{2-5} 
 & iWKB & $2.3011$ & $2.5225$ & $2.7004$\tabularnewline
\hline 
\hline 
\multirow{5}{*}{$0.0250$} & TCSA & $2.382$ & $2.659$ & $2.89$\tabularnewline
\cline{2-5} 
 & Airy & $2.3891$ & $2.6803$ & $2.9187$\tabularnewline
\cline{2-5} 
 & BS & $2.3816$ & $2.6590$ & $2.8813$\tabularnewline
\cline{2-5} 
 & WKB & $2.3791$ & $2.6583$ & $2.8812$\tabularnewline
\cline{2-5} 
 & iWKB & $2.3803$ & $2.6586$ & $2.8812$\tabularnewline
\hline 
\hline 
\multirow{5}{*}{$0.0375$} & TCSA & $2.497$ & $2.855$ & $3.15$\tabularnewline
\cline{2-5} 
 & Airy & $2.5099$ & $2.8915$ & $3.2039$\tabularnewline
\cline{2-5} 
 & BS & $2.4969$ & $2.8556$ & $3.1414$\tabularnewline
\cline{2-5} 
 & WKB & $2.4941$ & $2.8552$ & $3.1418$\tabularnewline
\cline{2-5} 
 & iWKB & $2.4954$ & $2.8553$ & $3.1417$\tabularnewline
\hline 
\hline 
\multirow{5}{*}{$0.05$} & TCSA & $2.597$ & $3.027$ & $3.37$\tabularnewline
\cline{2-5} 
 & Airy & $2.6177$ & $3.0800$ & $3.4584$\tabularnewline
\cline{2-5} 
 & BS & $2.5987$ & $3.0281$ & $3.3684$\tabularnewline
\cline{2-5} 
 & WKB & $2.5958$ & $3.0283$ & $3.3701$\tabularnewline
\cline{2-5} 
 & iWKB & $2.5971$ & $3.0282$ & $3.3698$\tabularnewline
\hline 
\hline 
\multirow{5}{*}{$0.1$} & TCSA & $2.933$ & $3.588$ & $4.11$\tabularnewline
\cline{2-5} 
 & Airy & $2.9805$ & $3.7143$ & $4.3151$\tabularnewline
\cline{2-5} 
 & BS & $2.9312$ & $3.5851$ & $4.0919$\tabularnewline
\cline{2-5} 
 & WKB & $2.9317$ & $3.5946$ & $4.1130$\tabularnewline
\cline{2-5} 
 & iWKB & $2.9320$ & $3.5933$ & $4.1114$\tabularnewline
\hline 
\hline 
\multirow{5}{*}{$0.2$} & TCSA & $3.447$ & $4.451$ & $5.22$\tabularnewline
\cline{2-5} 
 & Airy & $3.5565$ & $4.7213$ & $5.6750$\tabularnewline
\cline{2-5} 
 & BS & $3.4115$ & $4.3556$ & $5.0251$\tabularnewline
\cline{2-5} 
 & WKB & $3.4474$ & $4.4508$ & $5.2257$\tabularnewline
\cline{2-5} 
 & iWKB & $3.4426$ & $4.4455$ & $5.2209$\tabularnewline
\hline 
\hline 
\multirow{5}{*}{$0.4$} & TCSA & $4.22$ & $5.71$ & $6.85$\tabularnewline
\cline{2-5} 
 & Airy & $4.4707$ & $6.3198$ & $7.8337$\tabularnewline
\cline{2-5} 
 & BS & $3.8964$ & $4.8679$ & $4.9503$\tabularnewline
\cline{2-5} 
 & WKB & $4.2293$ & $5.7277$ & $6.8706$\tabularnewline
\cline{2-5} 
 & iWKB & $4.2074$ & $5.7102$ & $6.8559$\tabularnewline
\hline 
\end{tabular}
\par\end{centering}

\protect\caption{\label{tab:Meson-masses-ising}Meson masses in the Ising model. Theoretical
predictions are shown with $4$ digits accuracy, while for TCSA we
show the digits that can be reliably extracted (with the last digit
having an estimated precision of order $1$).}

\end{table}

\subsection{Three-state Potts model}

For the three-state Potts model we only consider the domain $h<0$,
since according to the discussion in subsection \ref{par:Mesonic-states-in}
there are no meson one-particle levels for $h>0$, and reading off
meson masses from two-particle states is difficult both for numerical
reasons (the spectrum is dense, so level identification is difficult)
and for theoretical reasons (extraction of masses with any precision
requires modeling the meson-meson scattering). In addition, as discussed
in subsection \ref{par:Baryon-masses-in}, there are no baryon states
for $h>0$.

\subsubsection{False vacuum}

\begin{figure}
\begin{centering}
\subfloat[$\tilde{h}=-0.005$]{\centering{}\includegraphics[scale=0.35]{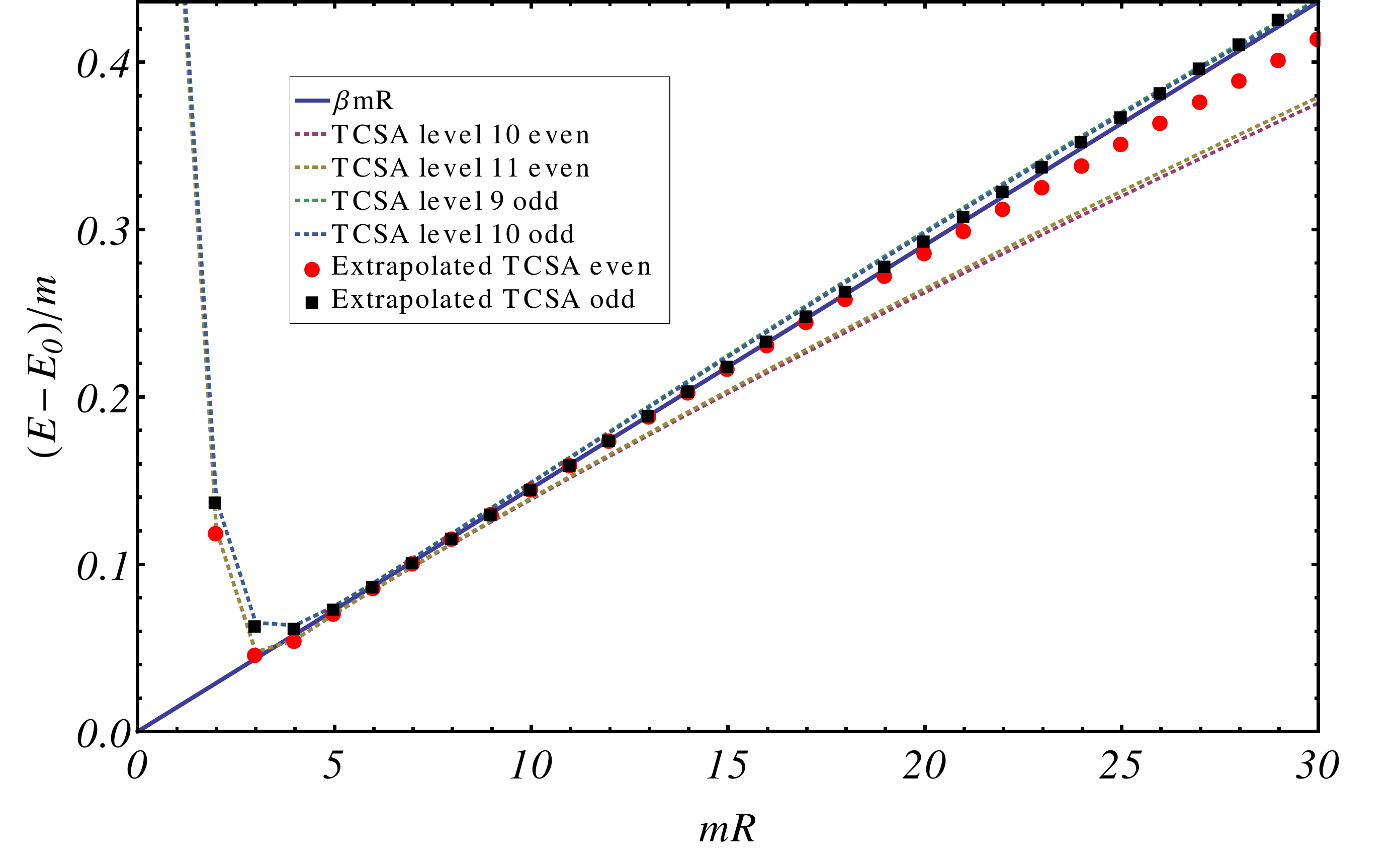}}\\
\subfloat[$\tilde{h}=-0.01$]{\centering{}\includegraphics[scale=0.35]{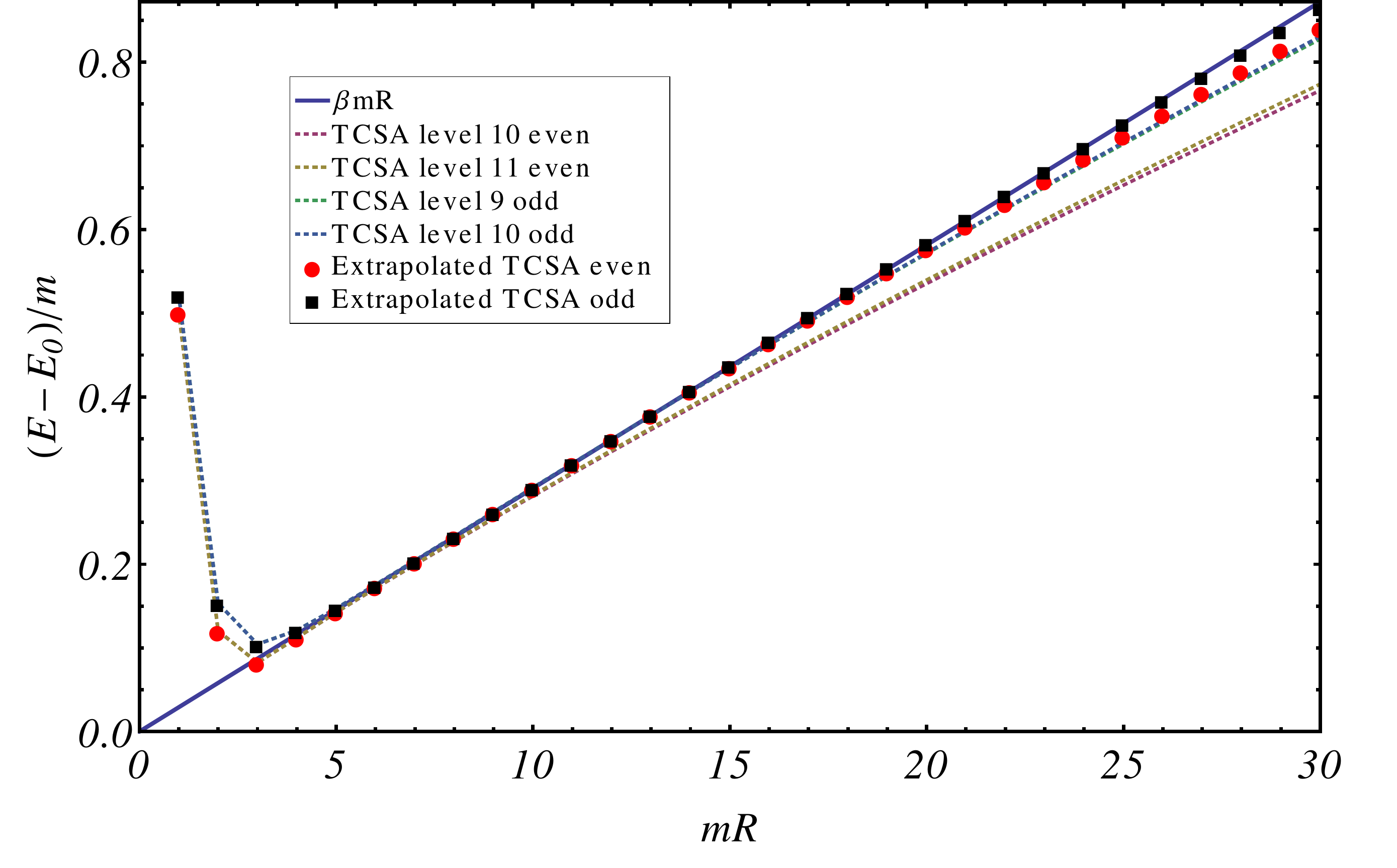}}\\

\par\end{centering}

\centering{}\subfloat[$\tilde{h}=-0.02$]{\centering{}\includegraphics[scale=0.35]{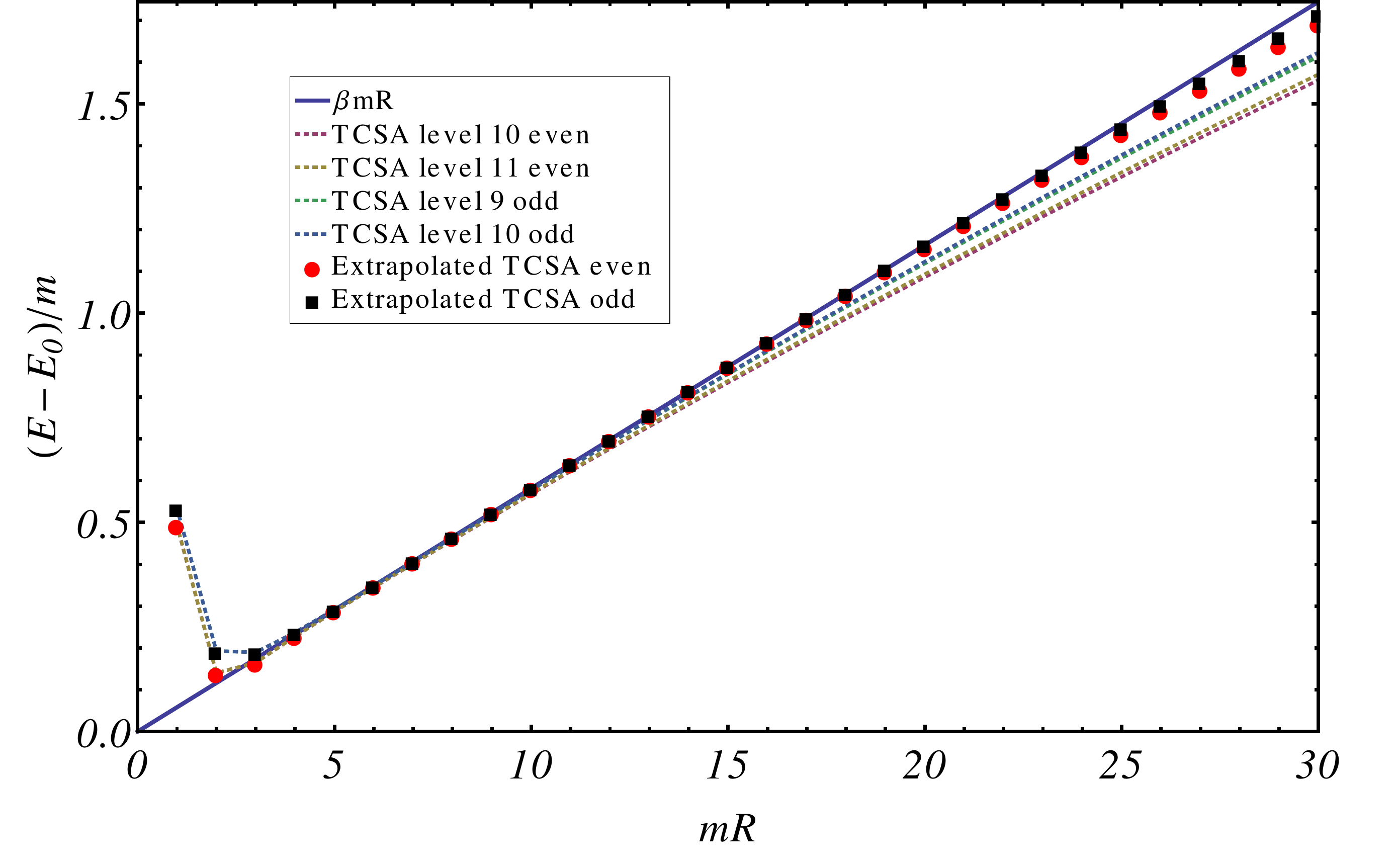}}\protect\caption{\label{fig:false_vacuum_Potts}Energy of the false vacuum for different
values of the magnetic field in the three state Potts model. Continuous
lines are the theoretical predictions from (\ref{eq:fv_endensity}),
dashed lines are data with coupling constant renormalization before
extrapolation for some values of the cut-off. Red dots are extrapolated
data in the $\mathcal{C}$-even sector, while black squares are extrapolated
data from the $\mathcal{C}$-odd sector. }
\end{figure}

The relative energy of the false vacuum against the volume is shown
in Figure \ref{fig:false_vacuum_Potts}. Notice that the extrapolation
is again very effective.

\subsubsection{Meson masses \label{sub:Meson-masses}}

\begin{figure}
\centering{}\includegraphics[scale=0.5]{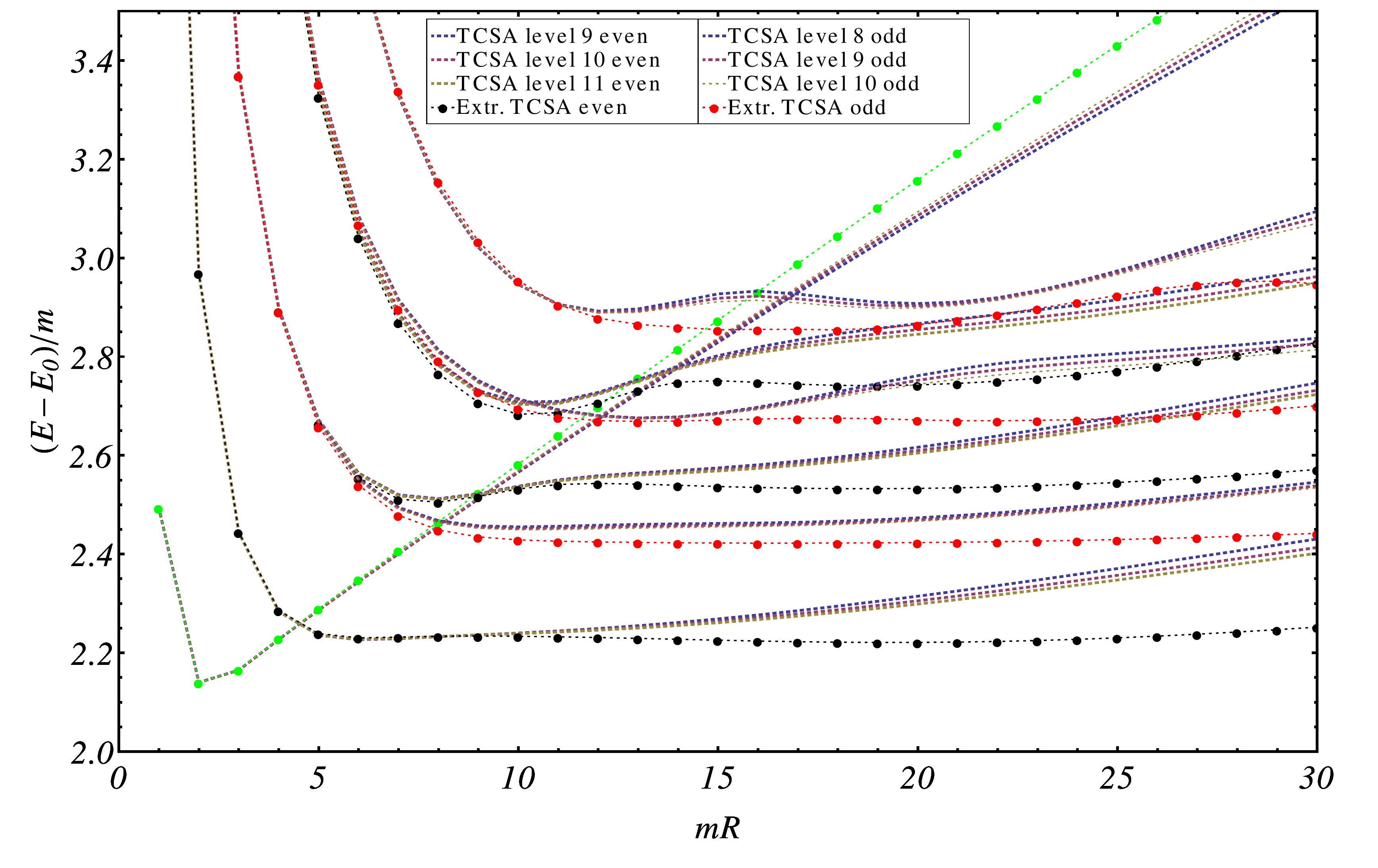}\protect\caption{\label{fig:Potts_mesons_shiftedFV}Effect of the extrapolation for
the first three mesonic state in the $\mathcal{C}$-even and odd sectors
with magnetic field $\tilde{h}=-0.02$. Dashed lines for data with
running coupling for some values of the cut off, while large dots
with dotted lines are the extrapolated data. We also plotted the energy
of the even sector false vacuum (both before and after extrapolation,
the latter marked with green) shifted up by $2m$ to demonstrate that
the wavy feature corresponds to the ``false meson'' resonance (meson
configuration over the false vacuum).}
\end{figure}

\begin{figure}
\centering{}\includegraphics[scale=0.6]{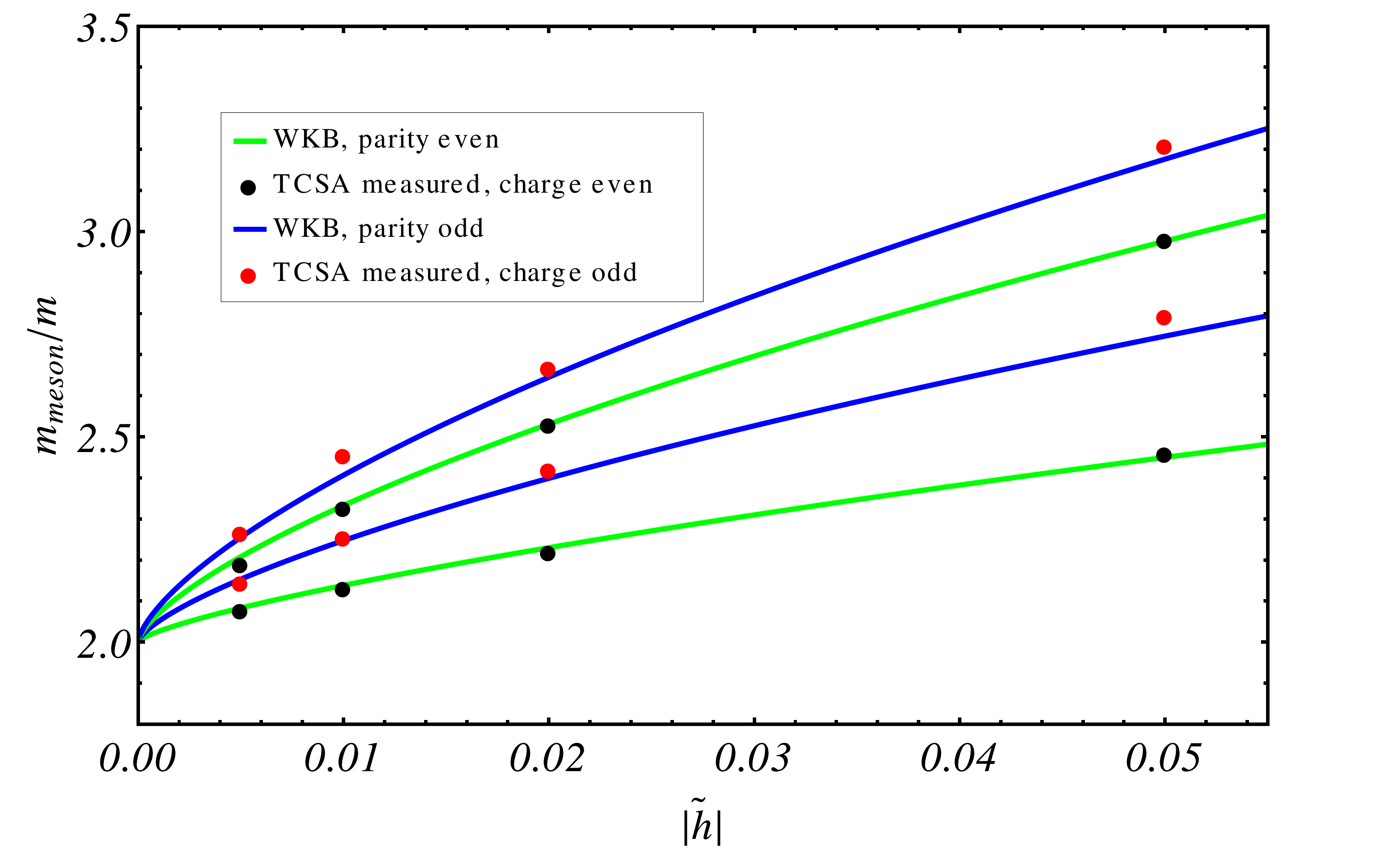}\protect\caption{\label{fig:Potts_meson_masses}Meson masses against the magnetic field
in the three state Potts model. Black dots show TCSA results from
the $\mathcal{C}$-even sector while red is for $\mathcal{C}$-odd.
The green and blue lines show the WKB predicted for parity even and
odd states. As expected, for mesons the two parities coincide.}
\end{figure}

In contrast to the Ising case here we can only use the method of extracting
the value at the flattest portion of curve to estimate meson masses,
due to the presence of the ``false meson'' resonance corresponding
to a kink-antikink bound state configuration starting and ending in
one of the false vacua. The ``wavy'' feature these resonance plateaus
\cite{Luscher:1991cf,Pozsgay:2006wb} introduce in the spectrum prevent
application of the exponential fit (\ref{eq:expfit}) to eliminate
finite size effects in the meson mass. This effect can be seen in
the plot \ref{fig:Potts_mesons_shiftedFV}, which also demonstrates
the efficiency of the numerical extrapolation procedure.
 The mesonic spectrum against the absolute values of the magnetic
field can be seen on figure \ref{fig:Potts_meson_masses}. The deviations
between the WKB prediction and the numerically determined masses are
typically of the order of a few times $10^{-3}$, except in a few
cases when a larger deviation of order $10^{-2}$ is observed. These
are cases when the flattest portion of the meson level contains a
level crossing with the false vacuum, which makes the truncation level
extrapolation less precise.

\subsubsection{Baryon masses}

\begin{figure}
\begin{centering}
\includegraphics[scale=0.6]{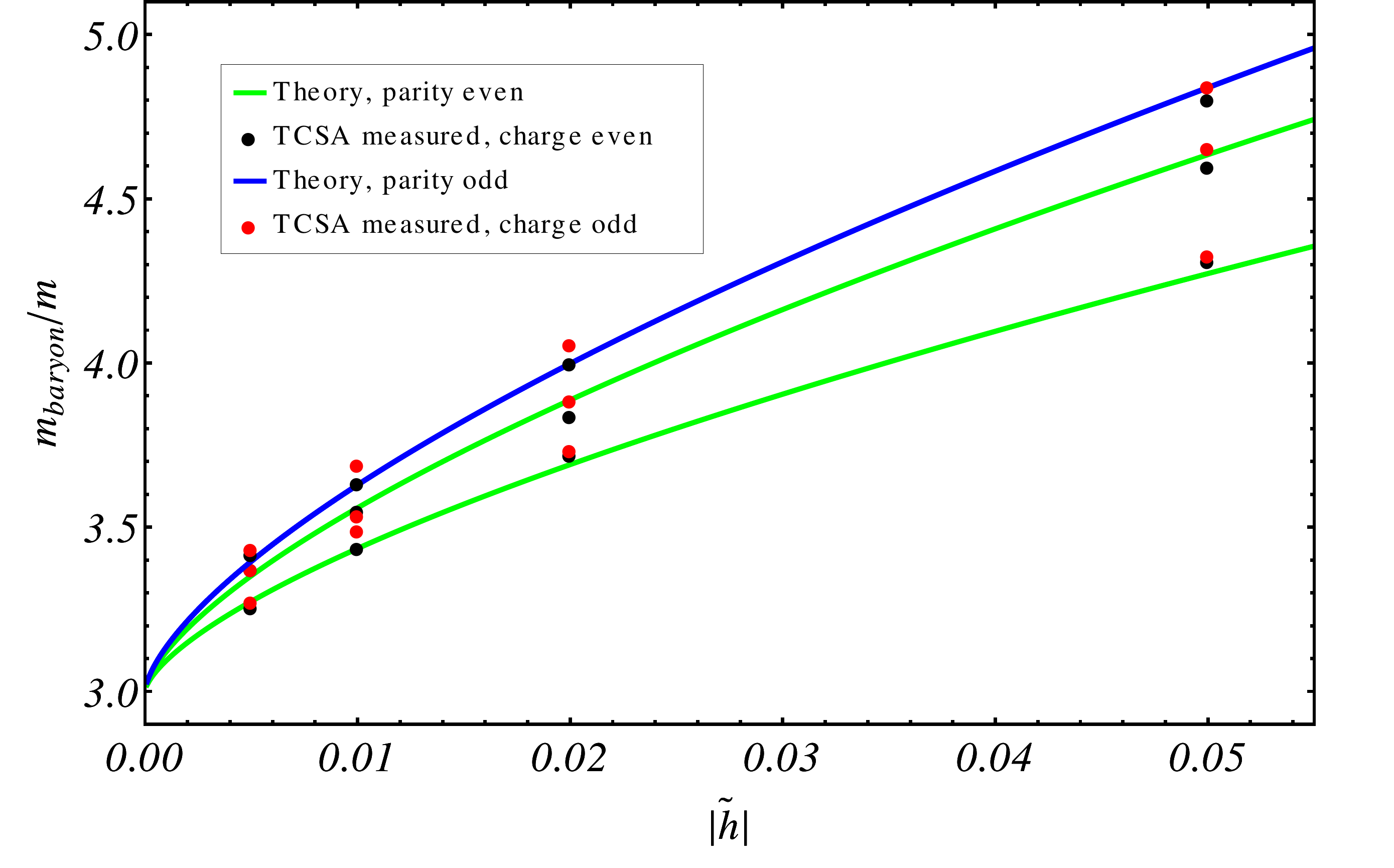}
\par\end{centering}

\protect\caption{\label{fig:Baryon_masses}The three lowest baryon masses against the
magnetic field in the three state Potts model. Black dots show TCSA
results from the $\mathcal{C}$-even sector while red is for $\mathcal{C}$-odd.
The green and blue lines show the predictions (\ref{eq:rutkevich_predictions},\ref{eq:rutkevich_epsilons})
for space parity even and odd states, respectively. It can be seen
that for each spatial parity there are two states, corresponding to
the $\mathcal{C}$-even/odd combinations of the baryon with its antiparticle.}
\end{figure}

The baryonic state{\small{}s} are in the higher part of the spectrum.
As a result, because of the many level crossings the state must be
carefully identified for each value of the volume and cut-off in order
to carry out the extrapolation. The masses are extracted as the value
of the extrapolated energy levels at its flattest point. As noted
in subsection \ref{par:Baryon-masses-in}, baryons and antibaryons
have the same spectra in infinite volume. In finite volume the eigenstates
are the charge conjugation ($\mathcal{C}$) even and odd combinations,
as can be seen from the results shown in Figure \ref{fig:Baryon_masses}.
The deviations between the theoretical prediction and the numerically
determined masses are typically of the order of a few times $10^{-3}$,
except in a few cases when a larger deviation of order a few times
$10^{-2}$ is observed, which in this case is mostly due to difficulties
of locating the level in the dense part of the spectrum.

\section{Conclusions\label{sec:Conclusion}}

In this work we investigated confinement in the $q$-state scaling
Potts field theory, for the cases $q=2$ (Ising) and $q=3$ (three-state
Potts). While these phenomena in the Ising model have been investigated
in numerous works \cite{FonZam2003,Rutkevich2005,FonZam2006,Rutkevich2009,Rutkevich2010}
since the seminal paper by McCoy and Wu \cite{McCoyWu1978}, resulting
in a very detailed understanding of the meson spectrum, in the case
of the three-state Potts the theoretical predictions are more recent,
especially regarding the baryon spectrum. Our method of choice was
the renormalization group improved truncated conformal space approach
(RG-TCSA), a Hamiltonian truncation method applied to perturbed conformal
field theories, since there is no alternative for the three-state
Potts model. 

Ising model was used both as a benchmark of the method, and as a way
of comparing the effectiveness of theoretical predictions. Our conclusion
was that semiclassical (WKB) quantization was efficient over all the
range of weak magnetic field, and therefore could be taken as reference
for the analysis of the Potts meson spectrum. Indeed, we could demonstrate
very good agreement between the WKB and the meson spectra predicted
in \cite{RutPottsMes2010}. In addition, we compared the recent predictions
the numerical results to recent predictions for the baryon spectrum
\cite{Rut2015}, and found complete agreement.

The present results lead to the conclusions that from a quantitative
phenomenological viewpoint, the meson spectrum of the the $q$-state
scaling Potts field theory in weak magnetic field is described by
WKB to a very high precision, while the baryon spectrum can be efficiently
modeled by the three-body quantum mechanical model introduced in \cite{Rut2015}.
These findings are expected to be relevant in future investigations
of the Potts field theory as a description of statistical systems.

\subsection*{Acknowledgments}

The authors are grateful to S. Rutkevich for sharing his unpublished
results in \cite{Rutkevich2010} and for discussions. G.T. also acknowledges
useful comments on the topic of RG-TCSA by S. Rychkov. This work was
supported by the Momentum grant LP2012-50/2014 of the Hungarian Academy
of Sciences.

\bibliographystyle{utphys}
\bibliography{magneticpotts}

\end{document}